\documentclass{article}

\usepackage[preprint,nonatbib]{neurips_2026}

\usepackage{graphicx}
\graphicspath{{./images/}}
\usepackage[utf8]{inputenc}
\usepackage[T1]{fontenc}
\usepackage{microtype}
\usepackage{hyperref}
\hypersetup{hidelinks}
\usepackage{url}
\usepackage{xurl}

\usepackage{amsfonts}
\usepackage{amssymb}
\usepackage{nicefrac}

\usepackage{booktabs}
\usepackage{makecell}
\usepackage{longtable}
\usepackage{tabularx}
\usepackage{array}
\usepackage{enumitem}
\usepackage{float}
\usepackage{placeins}
\usepackage{pdflscape}
\usepackage{xcolor}

\usepackage[
  style=authoryear,
  backend=biber,
  natbib=true
]{biblatex}
\newcolumntype{Y}{>{\raggedright\arraybackslash}X}
\newcolumntype{L}[1]{>{\raggedright\arraybackslash}p{#1}}

\renewcommand{\arraystretch}{1.2}

\setlist[itemize]{nosep,leftmargin=*,topsep=0pt,label=\textbullet}

\usepackage[labelfont=bf]{caption}
\selectfont
\makeatletter
\newcounter{vignette}
\newcommand{\listvignettename}{List of C2 Vignettes}
\newcommand{\listofvignettes}{%
  \section*{\listvignettename}%
  \@starttoc{lov}%
}
\newcommand{\vignettetitle}[1]{%
  \refstepcounter{vignette}%
  \addcontentsline{lov}{figure}{\protect\numberline{\thevignette}#1}%
  \textit{\textbf{C2 vignette \thevignette: #1}}%
}
\makeatother

\usepackage[acronym]{glossaries}
\makeglossaries

\title{Testing and Evaluation of Agentic AI Systems In Military Command and Control}

\author{
{\normalfont
\begin{tabular}{ccc}
\bfseries Ulysse Richard$^{1}$\thanks{Corresponding author: \texttt{ulysse.richard@arcadiaimpact.org}} & \bfseries Heather Frase$^{2}$ & \bfseries Sarah Cao$^{1,3}$ \\[0.7em]
\bfseries Di Cooke$^{1,4}$ & \bfseries Sebastian Kwon$^{1,5}$ & \bfseries Adrianna Tan$^{1,6}$
\end{tabular}
}\\[1.8em]
{\normalfont $^{1}$ Arcadia Impact, AI Governance Taskforce}\\
{\normalfont $^{2}$ Veraitech}\\
{\normalfont $^{3}$ University of Oxford}\\
{\normalfont $^{4}$ King's College London}\\
{\normalfont $^{5}$ Atlantic Council}\\
{\normalfont $^{6}$ Future Ethics Lab}\\
}

\begin{document}

\maketitle
\begin{center}
\vspace{-2.2em}
August 2026
\vspace{0.8em}
\end{center}

\begin{abstract}

Agentic AI systems are being procured for military command and control (C2) under public commitments to rigorous testing and human oversight. Whether such commitments can be discharged depends on their supporting assurance case, which requires three elements: claims specifying the conditions for acceptability, evidence bearing on those claims, and an argument connecting the two. Through a structured review of 240 documented Testing and Evaluation (T\&E) practices, spanning eight evaluation dimensions and three lifecycle stages, we identify eight assumptions that established methods make about their test article, grouped into four clusters: system specifiability, stability, composability, and supervisability. Agentic properties weaken all eight assumptions. This erosion affects the argument connecting evidence to claims, not the claims or evidence themselves. As a result, test results may satisfy process requirements, but they do not warrant the inference from tested to fielded behavior.

We derive ten assurance claims for the first three assumption clusters and assess whether current and emerging methods can address each, mapping operational consequences through five C2 scenarios. Supervisability is identified but not assessed here, since evidencing it depends on system stability results and human factors T\&E methods beyond the present scope. The documented record does not support broad claims about system-level behavior, but narrower claims remain recoverable, contingent on mature methods: bounded mission envelopes, trajectory-grounded correctness, executable runtime constraints, and characterized run-to-run variance. Part of the evidentiary burden shifts into deployment, making the determination to field a continuing act. Where evidence cannot be generated, the residual uncertainty can be governed through defined expiry conditions and assigned ownership.
\end{abstract}

\newpage
\raggedbottom

\phantomsection
\section*{Executive Summary}
\addcontentsline{toc}{section}{Executive Summary}

\begin{figure}[H]
  \centering
  \includegraphics[width=\textwidth]{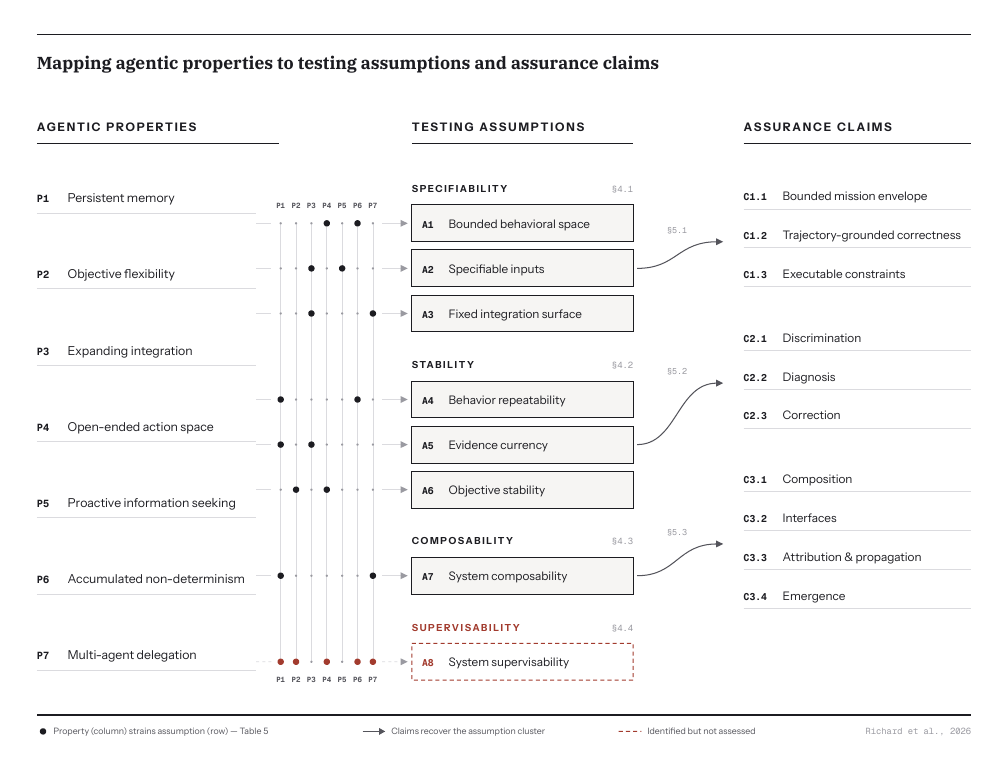}
  \caption{Mapping agentic properties to testing assumptions and assurance claims.}
  \label{fig:summary}
\end{figure}

This research paper examines how much confidence current Testing and Evaluation (T\&E) methods can justify for agentic AI systems in command and control (C2), and how remaining uncertainty should be handled in decisions to field them. Figure~\ref{fig:summary} serves as a visual summary of this paper.

\textbf{Why examine this challenge now?}

\begin{itemize}
  \item Agentic AI capabilities are being procured for C2 roles; the basis for assurance commitments warrants attention from military, industry, and policy communities.
\end{itemize}

\textbf{Where do established T\&E methods fall short?}

\begin{itemize}
  \item In agentic systems, the unit of behavior is a trajectory encompassing tool use, memory, and delegation, rather than a discrete output.
  \item Certain system elements emerge only at runtime. Agents may select sources and tools, or subagents can be added after certification, creating differences between tested and deployed configurations.
  \item The test article is inherently dynamic; a system that accumulates state\footnote{In this paper, ``state'' primarily refers to an evolving memory state that may take the form of a text buffer, key-value store, vector database, graph structure, or any hybrid representation, as defined in \citet{hu_memory_2025}.} is not the same as when it was originally characterized. Evidence can age without explicit updates, and objectives may shift under operational pressures.
  \item Assemblies introduce challenges absent in isolated components. Emergent behaviors arise within the assembly, complicating attribution, and valid outputs at the component level may conflict when composed.
  \item Supervision is itself strained. Opacity, adaptation, and delegation erode the operator's mental model, reduce its intervention window, and increase supervisory load. Human oversight cannot be assumed from the mere presence of an operator.
\end{itemize}

\textbf{How can justified confidence be recovered?}

\begin{itemize}
  \item Confidence should be structured as an assurance case comprising claims, evidence, and an argument that connects the two. For agentic systems, the argument is the weak link, because test-condition evidence no longer corresponds to fielded behavior in the ways established methods assume.
  \item While broad assurance claims remain unsubstantiated, more narrowly scoped claims remain recoverable contingent on mature methods, including bounded mission envelopes, trajectory-grounded correctness, executable runtime constraints, and characterized run-to-run variance.
  \item Part of the evidentiary burden shifts into deployment (behavioral monitoring, re-baselining, goal-drift detection, staged fielding), making the determination to field a continuing act. What cannot be evidenced must be governed, borrowing from assurance regimes in software, aviation, nuclear energy, medicine, finance, and autonomous driving.
\end{itemize}

\textbf{Key contributions of this paper}

\begin{itemize}
  \item Seven agentic properties, mapped to a generic architecture and five illustrative C2 scenarios (\S2).
  \item Eight assumptions underlying established T\&E practice that agentic properties strain, drawn from 240 documented practices (\S3, \S4).\footnote{The dataset is available from the corresponding author on reasonable request.}
  \item Ten assurance claims covering specifiability, stability, and composability, with candidate methods and their evidentiary limits assessed against each (\S5). Supervisability is treated only at the interface through which human direction enters the system, for the reasons given in \S1.3.
\end{itemize}

\begin{table}[H]
\small
\centering
\setlength{\tabcolsep}{6pt}
\renewcommand{\arraystretch}{1.25}
\begin{tabularx}{\textwidth}{|L{0.30\textwidth}|X|L{0.17\textwidth}|}
\hline
\textbf{Finding} & \textbf{Implication} & \textbf{Where addressed} \\
\hline
Established T\&E methods rest on assumptions about specifiability, stability, composability, and supervisability that agentic systems weaken by design. & Test evidence can satisfy process requirements while the inference to fielded behavior remains insufficient; the connecting argument needs direct scrutiny. & \S4 \\
\hline
Assurance claims are recoverable in a narrower form. & An assurance case decomposed into narrow claims with declared residual uncertainty can be supported by the documented record, whereas a broad system-level claim cannot. & \S5.1-5.3 \\
\hline
Some evidence can only be generated in service. & The determination to field becomes a continuing act under governance that time-bounds the validity of evidence. & \S5.2 \\
\hline
\end{tabularx}
\caption{Summary of findings and implications}
\label{tab:headline-findings}
\end{table}

\newpage

\tableofcontents

\newpage
\phantomsection
\addcontentsline{toc}{section}{Lists of Tables, Figures and C2 Vignettes}
\listoftables
\listoffigures
\listofvignettes
\newpage

\section*{Contribution Statement}

\textbf{Conceptualization, Methodology:} Ulysse Richard, Heather Frase \\
\textbf{Investigation:} Ulysse Richard, Sarah Cao, Di Cooke, Sebastian Kwon, Adrianna Tan

\textbf{Writing -- Original Draft:} \\
\vspace{0.2cm}
\begin{tabular}{@{}ll@{}}
$\quad$ Ulysse Richard & \S1, \S2, \S3, \S4, \S5, \S6 \\
$\quad$ Di Cooke       & \S5.1, \S5.2 \\
$\quad$ Sebastian Kwon & \S5.3 \\
$\quad$ Sarah Cao      & \S2.2.3, \S4, \S5.3 \\
$\quad$ Adrianna Tan   & \S5.2 \\
\end{tabular}

\textbf{Writing -- \LaTeX:} \\
\vspace{0.2cm}
$\quad$ Ulysse Richard

\textbf{Writing -- Review \& Editing, Supervision:} \\
\vspace{0.2cm}
$\quad$ Ulysse Richard, Heather Frase

\textbf{Project Administration:} Ulysse Richard

\section*{Acknowledgements}

The authors thank Ben R. Smith and Francesca Gomez for their feedback and project management support, and Virgile Richard for reviewing the manuscript. The paper draws on semi-structured expert interviews; we are indebted to all interviewees for their time and candor, as well as to the many others whose conversations informed this final product. Any remaining errors are our own.

This research was conducted as part of the AI Governance Taskforce at Arcadia Impact, Summer 2026 cohort.


\section*{AI use statement}

Large language models were used in the preparation of this paper. During the literature review, language models helped scan scientific databases, conduct exploratory reviews and syntheses, and perform supplementary searches to identify relevant articles not surfaced by structured queries. During analysis, AI tools helped organize extracted data into tables and graphs to support the mapping exercise. During writing, they were used for grammar and spelling, prose review and editing, figure preparation, and LaTeX formatting. The expert interviews were conducted by human researchers. All substantive analytical claims are the authors' own, and any AI-assisted text and outputs were reviewed and verified by the authors, who take full responsibility for the content of the paper.

\newpage


\section{Introduction}

Assurance commitments for agentic AI systems appear in program announcements, defense Testing and Evaluation (T\&E) strategies, and policy directives. These commitments cover matters such as rigorous testing throughout development and fielding, the retention of human judgment over consequential decisions, and continuous evaluation throughout the system's lifecycle rather than ending at acceptance. Each commitment requires supporting evidence. For agentic systems, methods for generating such evidence fall into three categories: those currently available and usable; those available but not yet validated for this class of system; and those for which no candidate method has been identified. A structural challenge is that both emerging and established T\&E methods rely on assumptions about system behavior that agentic systems weaken by design. This paper identifies affected assumptions, evaluates the value and limitations of current methods, and identifies potential methods and governance mechanisms to address remaining gaps.

In June 2026, the United States Department of War launched ``Agent Network,'' a capability that will leverage artificial intelligence (AI) agents to continuously scan defense intelligence and operational systems and present commanders with options \citep{noauthor_dow_2026}. The announcement commits to subjecting the capability to ``rigorous testing, operational evaluation and oversight'' throughout development and fielding. As comparable capabilities are being developed by multiple militaries or may soon be, the basis for such commitments warrants collective attention from military, industry, and policy communities, both nationally and internationally.

Battle management and decision support are command and control (C2) activities, and the Department describes Agent Network as building on prior C2 work. C2 is a demanding setting for these systems. It combines interconnected sensor networks, incomplete information, adversary manipulation, and real-time operational consequence, and its outputs pass through human operators whose reliance on the system is itself a performance variable.\footnote{\S2.2 develops the C2 setting and the definition we adopt for it.}

This paper is organized around two questions. How much confidence can current evaluation methods justify for agentic AI systems in command and control? How should residual uncertainty be addressed or accounted for in decisions to field these systems?

\subsection{Locating the challenge}

Agentic AI systems create T\&E challenges across three nested levels. The \textbf{model} refers to the frontier AI system that performs interpretation, reasoning, and action selection. The \textbf{agent} comprises the model and its surrounding scaffolding, including memory, tools, and an orchestration loop. The \textbf{deployed system} consists of an assembly of agents and operators, along with their organizational context.

\textbf{At the model level}, evaluating frontier AI systems remains an open problem for their developers, as illustrated by the disclosure of failures in their evaluation programs in July 2026 \citep{openai_openai_2026,anthropic_investigating_2026}. Though these incidents involve cyber capabilities, they are symptoms of a broader diagnosis: T\&E practice remains largely ad hoc and lacks grounding in standards or scientific engineering principles \citep{department_for_science_innovation__technology_frontier_2025}. Because the internal mechanisms of frontier AI systems are not interpretable to their developers, evaluation strategies primarily rely on observing behavior in response to selected inputs. This observation is inherently incomplete, allowing capabilities and failure modes to emerge after deployment. Agents built on such models inherit these limitations, and while scaffolding may constrain their effects, it does not eliminate them.

\textbf{At the agent level}, the scaffolding introduces complexities not present in the model alone. Established T\&E methods do not require that every behavior be listed in advance. They do require that the space of possible behavior be partitionable, parameterizable, or otherwise bounded well enough for a sample to stand for the whole. Agentic systems weaken that condition. They pursue objectives autonomously, construct their own information environment by selecting sources to query, act on that environment using tools, and retain context across runs and sessions. The properties that make agents useful are often the same properties that make them difficult to evaluate \citep{anthropic_demystifying_2026}. For example, variation that testers have characterized statistically for a single model output behaves differently along a chain of steps, where each output becomes the next input, altering the implications of repeated trials. Additionally, an agent that retains state is not the same test article\footnote{In testing, a test article is the specific component, system, or system-of-systems that is undergoing evaluation.} from one week to the next, even in the absence of a declared change to the system.

\textbf{At the deployed system leve}l, agents are assembled with other agents and with human operators, introducing additional complexity. Certain behaviors emerge only in the assembly and resist attribution to individual components. Moreover, fielding an agentic system changes who decides what, who exchanges information with whom, and who knows what, thereby affecting the performance of the human-machine team without altering the technology. Agent-to-agent delegation also raises the tempo and scale of activity, reducing the time between a failure occurring and its consequences reaching the operational picture. These effects are not observable when testing models or single agents in isolation, raising questions about the validity of component-level evidence for the assembled system.

\subsection{The structure of justified confidence}

In AI assurance, justified confidence in a fielded system is established through an assurance case comprising three elements \citep{tate_framework_2016}:

\begin{itemize}
  \item \textbf{Claims}, the \textit{specification}, state what must hold for the system to be judged acceptable for its intended use.
  \item \textbf{Evidence}, the \textit{demonstration}, is what has been observed of the system and bears on those claims.
  \item An \textbf{argument} connects the evidence to the claim, explaining why evidence gathered under specific conditions supports a claim about the system as fielded. All three elements are necessary.
\end{itemize}

Claims without evidence produce confidence with no basis. Evidence without claims produces unfocused testing and data that fulfill process requirements without informing judgment. Claims and evidence without a sound connecting argument create the appearance of assurance while leaving the inference unexamined. This third case is the principal concern of this paper.

Test and Evaluation is the disciplined collection and analysis of evidence about the behavior and performance of systems, used to understand them, improve them, and assure that they are safe and fit for purpose \citep{uk_ministry_of_defence_test_2026}. Within assurance, we distinguish three further roles:

\begin{itemize}
  \item \textbf{Testing} generates data about system behavior under stated conditions.
  \item \textbf{Evaluation} applies judgment to that data against stated criteria.
  \item The \textbf{determination} to field weighs that judgment alongside operational need, cost, and schedule. It rests with an authority that is typically external to the test organization. This paper focuses on the contributions of testing and evaluation to the determination, rather than the determination itself.
\end{itemize}

T\&E evidence is collected from a particular article, under particular conditions, at particular sampled points, and using particular measures. Such evidence warrants a claim about the fielded system only to the extent that each of these aspects corresponds to operational reality, and the strength of that warrant depends on the quality of the argument for that correspondence. Agentic AI systems simultaneously weaken several of these correspondences across the three levels described above.

\subsection{Scope and approach}

This paper examines the challenges that the unique properties of agentic AI systems pose to the T\&E process, with a focus on the C2 context. Taking established T\&E practice as a starting point, we identify where agentic properties require new methods or governance mechanisms. We also consider governance mechanisms that determine when evidence is collected and when its validity expires, as these mechanisms are inseparable from the evidence question for systems that change during service.

Narrow rule-based systems, classical machine learning, and single-turn generative AI without retained context are excluded from the scope of this analysis. \S4 identifies the assumptions that agentic properties place under strain, and \S5 assesses what current methods can establish against the specifiability, stability, and composability clusters, and what would be required to close the remaining gaps.

\S5 addresses three of the four assumption clusters, namely specifiability, stability, and composability. Supervisability, identified in \S4.4, is carried forward only through the interface by which human direction enters the system. It concerns the operator-system pair rather than the system. Evidencing supervisability rests on human-subjects methods that are not included in the T\&E corpus reviewed here. The assessment of supervisability is further contingent on the stability results in \S5.2, since operator reliance can only be calibrated against a defined performance baseline. We return to it in \S6 as a research priority.

T\&E is assessed against eight dimensions: functional performance (D1), non-functional performance (D2), robustness (D3), behavioral stability (D4), differential performance (D5), security (D6), safety (D7), and human-machine teaming (D8), each defined in Appendix~\ref{app:dimensions}. These dimensions were examined across three AI lifecycle stages: component-level characterization, system-level integration testing, and post-deployment monitoring. \S4 and \S5 are organized by the assumptions underlying the methods for these dimensions, rather than dimension by dimension. The eight dimensions guide the extraction of practices in Section 3, while the eight assumptions synthesize insights across those dimensions to provide the analytical framework for Sections 4 and 5. A single assumption typically draws on methods that span several dimensions.

\section{Definitions and Concepts}

\subsection{Agentic AI}

Agentic AI systems are characterized by a set of high-level faculties, functions, and architectural components.

\textbf{High-level faculties.} Agency is understood as a continuum, with systems exhibiting varying degrees of agentic behavior based on specific properties. A system demonstrates agency when it pursues assigned goals without explicit procedural instructions, acts upon its environment independently of human mediation, and adapts to unforeseen circumstances. Four defining characteristics of agentic systems are identified, drawing on \citet{wooldridge_intelligent_1995} weak notion of agency and related concepts in recent literature (Table~\ref{tab:faculties}).

\begin{table}[H]
\footnotesize
\centering
\setlength{\tabcolsep}{5pt}
\renewcommand{\arraystretch}{1.25}
\begin{tabularx}{\textwidth}{|L{0.13\textwidth}|X|X|}
\hline
\textbf{Faculties} & \textbf{Definition} & \textbf{Related terms in the literature} \\
\hline
\textbf{Autonomy} & The system operates without direct human intervention and controls its own actions and internal state. & Independent execution \citep{shavit_practices_2024}; directness of impact \citep{chan_harms_2023,kraprayoon_ai_2025}; degree of user supervision \citep{kapoor_ai_2024} \\
\hline
\textbf{Reactivity} & The system perceives its environment and responds to changes within it, including novel or unanticipated circumstances. & Adaptability \citep{shavit_practices_2024,kraprayoon_ai_2025}; environments subject to unexpected change \citep{kapoor_ai_2024} \\
\hline
\textbf{Pro-activeness} & The system takes the initiative to pursue an objective rather than acting only in response to its environment. An operator supplies the objective without a specification of how it is to be accomplished, and the objective may be complex or extend over a long horizon. & Underspecification and goal-directedness \citep{chan_harms_2023}; goal complexity \citep{shavit_practices_2024}; pursuit of goals without instruction on how to pursue them \citep{kapoor_ai_2024} \\
\hline
\textbf{Social ability} & The system communicates with humans and other systems using a formal language, allocates tasks among additional instances of itself, and functions in environments with multiple stakeholders. & Multi-agent collaboration, including delegation to subagents \citep{kraprayoon_ai_2025}; collaboration \citep{rickli_international_2026}; others treat multiple stakeholders as an attribute of the environment \citep{kapoor_ai_2024,shavit_practices_2024} \\
\hline
\end{tabularx}
\caption{Faculties of agentic systems and related terms in the literature}
\label{tab:faculties}
\end{table}

\begin{figure}[H]
  \centering
  \includegraphics[width=\textwidth]{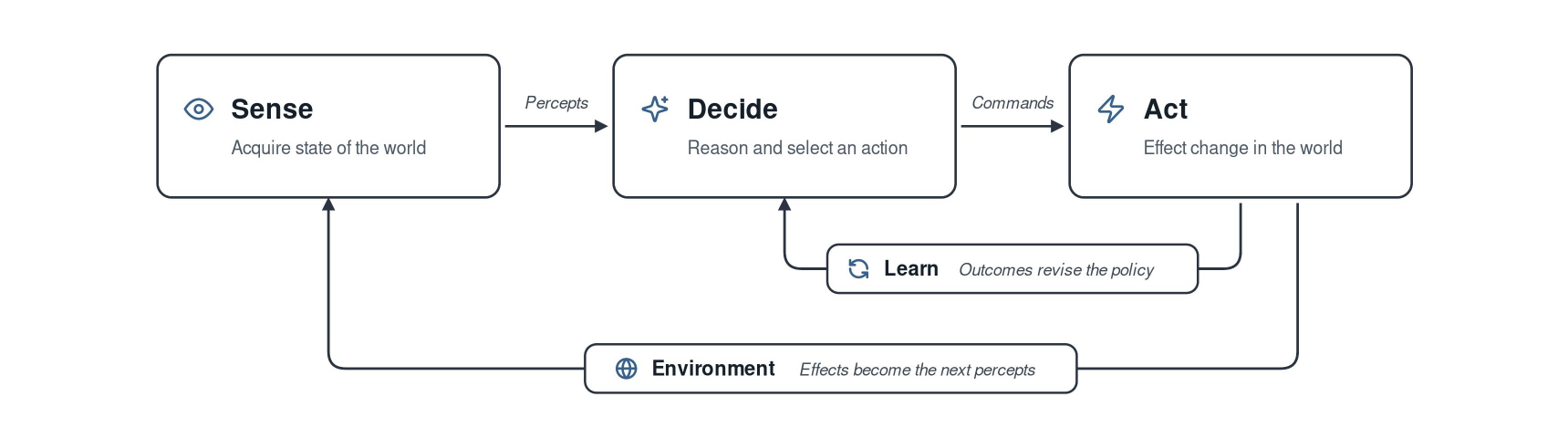}
  \caption{Functions of an agentic system}
  \label{fig:functions}
\end{figure}

\textbf{Functions.} At the functional level, AI agents are understood to perceive their environment, make decisions, act through actuators, and learn from their experiences \citep{russell_artificial_2022}. Disciplinary perspectives partition these functions in various ways. In human-machine teaming, automation is categorized as information acquisition, information analysis, decision and action selection, and action implementation \citep{parasuraman_model_2000}. Cognitive architectures identify a comparable set of core abilities, including perception, attention, action selection, memory, learning, and reasoning \citep{kotseruba_40_2020,laird_standard_2017}. Notably, learning is treated as a discrete function, distinct from task execution \citep{russell_artificial_2022,kotseruba_40_2020}. Figure~\ref{fig:functions} organizes these functions in a system view.

\begin{figure}[H]
  \centering
  \includegraphics[width=\textwidth]{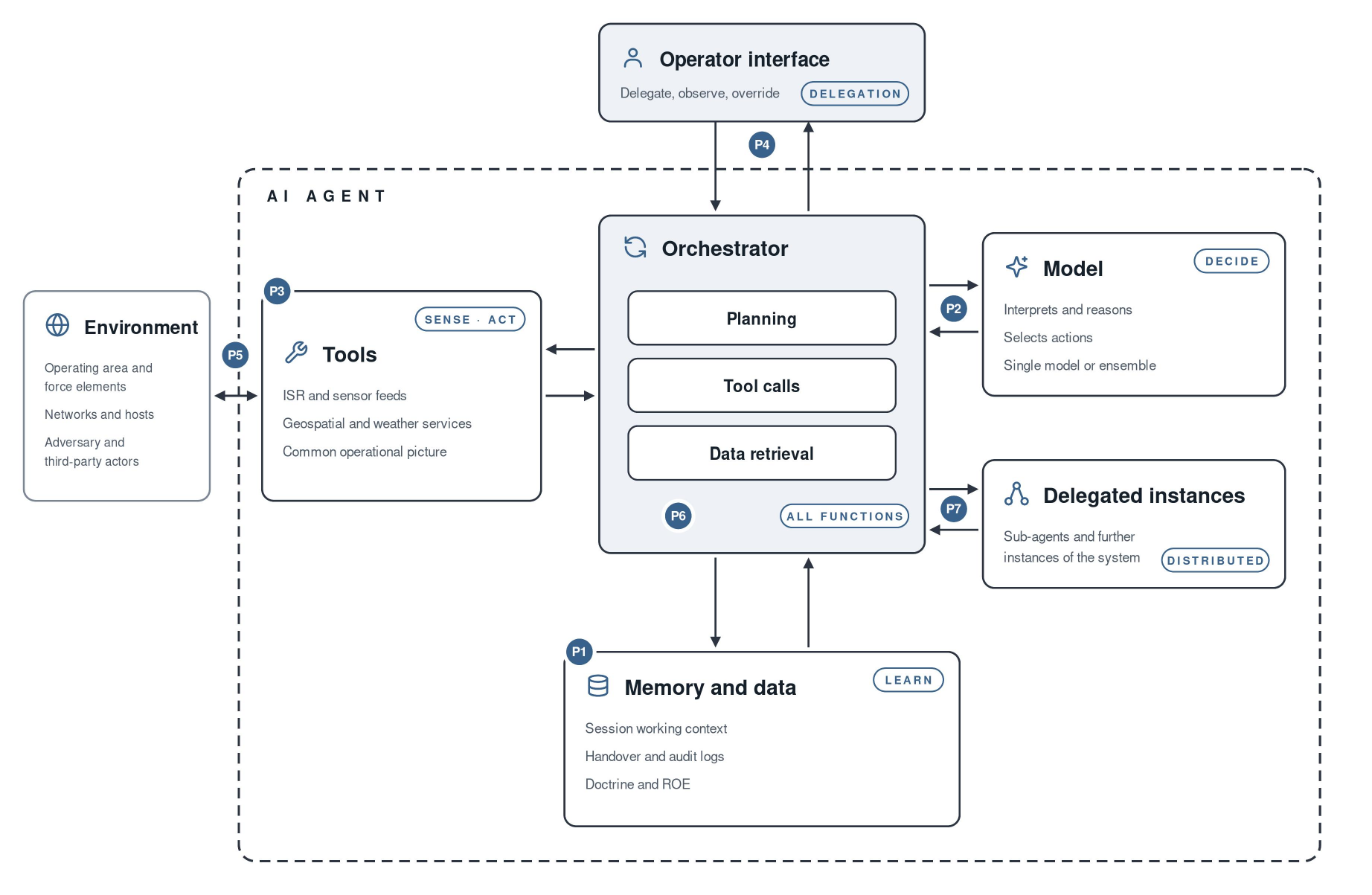}
  \caption{Generic agentic system architecture, with properties P1-P7 mapped to components}
  \label{fig:architecture}
\end{figure}

\textbf{Architecture.} Functions are realized through hardware and/or software components that collectively constitute the system. This component-based perspective determines the observable aspects of system operation. Figure~\ref{fig:architecture} presents a generic agentic architecture drawn from the technical and policy literature on agentic AI and from cognate systems in adjacent industries. It should be noted that it is illustrative and does not assert that a deployed command and control system will take this form.

The following reference set broadly outlines the role of each component within a command and control context:

\begin{itemize}
  \item \textbf{Operator interface}. Channel through which intent is delegated and recommendations are returned, and through which an operator observes and overrides
  \item \textbf{Orchestrator}. Control loop that sequences model calls, decomposes tasks, and delegates to further instances
  \item \textbf{Model}. Foundation model or models performing interpretation, reasoning, and action selection
  \item \textbf{Memory and data}. Working context within a session, state retained across sessions and operator handovers, and the corpora, indices, and retrieval policy from which the system draws.
  \item \textbf{Tools}. The declared set of callable capabilities and the channel through which they are invoked, including sensors, external services, and connected command and control systems.
  \item \textbf{Environment}. Operational setting is the system acts within and draws from, including physical and digital networks and actors, friendly or adversary.
  \item \textbf{Delegated instances}. subagents and further instances of the system to which the orchestrator assigns decomposed tasks.
\end{itemize}

\textbf{Properties.} The following seven properties in Table~\ref{tab:properties} operationalize the defining features of agentic AI systems and their implications for testing and evaluation (T\&E). Figure~\ref{fig:architecture} illustrates their possible location within a generic architecture. Several of these properties bear directly on command and control, including persistent memory across watch rotations and proactive information seeking against a contested picture, both of which recur in the illustrative scenarios of \S2.2.

\begin{table}[H]
\footnotesize
\centering
\setlength{\tabcolsep}{4pt}
\renewcommand{\arraystretch}{1.25}
\begin{tabularx}{\textwidth}{|L{0.025\textwidth}|L{0.135\textwidth}|X|L{0.085\textwidth}|L{0.135\textwidth}|}
\hline
\textbf{\#} & \textbf{Property} & \textbf{Description} & \textbf{Function} & \textbf{Component} \\
\hline
P1 & Persistent memory & The system preserves contextual information across sessions and operator transitions. As a result, a previously characterized instance may differ from its later operational state. & Learn & Memory and data \\
\hline
P2 & Objective flexibility & The system adjusts its priorities in response to contextual changes without explicit instructions. New goals may diverge from the initially assigned objectives. & Decide & Model; orchestrator \\
\hline
P3 & Expanding integration\footnotemark{} & The system functions across an expanding set of platforms, databases, and tools, and this set can grow beyond its certified configuration after deployment. & Act & Tools \\
\hline
P4 & Open-ended action space & New actions arise from goal-directed reasoning and cannot be exhaustively specified during design. Authorization boundaries frequently remain ambiguous when high-level goals are delegated. & Decide, Act & Orchestrator; tools; operator interface \\
\hline
P5 & Proactive information seeking & The system independently determines which sources to query, thereby shaping its own information environment. & Sense & Tools, memory, and data \\
\hline
P6 & Accumulated non-determinism & As each step informs the next, variance accumulates along the trajectory rather than remaining independent across steps. At operational chain lengths, outputs may differ substantially even from identical initial inputs, so that a reproducible test case yields a distribution of trajectories rather than a single expected result. & All, iterated & The loop from model through tools and memory and back \\
\hline
P7 & Multi-agent coordination and delegation & An orchestrator delegates tasks to subagents via unstructured outputs, such as natural-language interfaces, so that one agent's output becomes another's input. & All, distributed & The channel between orchestrator and delegated instances \\
\hline
\end{tabularx}
\caption{Properties of agentic systems, with associated functions and architectural components}
\label{tab:properties}
\end{table}
\footnotetext{Expanding integration refers to growth in the set of platforms, tools, and subagents the system can invoke, not to capability as such. Its consequences include a larger reachable action space, which compounds the open-ended action problem in P4, a wider vulnerability surface, and a broader boundary for testing to characterize.}

\subsection{Command and control}

\subsubsection{What is command and control?}

\textbf{Definition}. Command and control is the exercise of authority and direction by a properly designated commander over assigned and attached forces in the accomplishment of the mission \citep{department_of_defense_department_2010}. Both functions are performed through an arrangement of personnel, equipment, communications, facilities, and procedures employed by a commander in planning, directing, coordinating, and controlling forces and operations to accomplish the mission. Two features are critical for evaluation. First, it distinguishes command (setting intent and initial conditions) from control (modifying conditions as situations evolve), providing a boundary that can be located in a system where authority is delegated. Second, the account is agnostic regarding who performs each function or how. This makes it possible to describe a change in which entity performs a given function without first resolving whether doctrine permits that change.

\textbf{Cycle models and their limits}. The functional view of agents (sensing, deciding, acting, and learning) closely parallels command theory. Boyd's OODA loop (observe, orient, decide, act) shows that faster cycling than an opponent yields an advantage \citep{boyd_essence_1996,osinga_science_2007}. In this framework, orientation is shaped by experience and context, which in turn influence perception and action. The overlap between these functions makes the OODA loop a natural entry point for considering agent roles in command. However, Boyd's formulation is broad and admits multiple interpretations, a limitation the NATO C2-Cycle was developed in part to address by decomposing the sequence into collecting, decision-making, and effecting, organized around a connecting function (\citealp{van_rijn_ai_2025, nato_allied_command_transformation_nato_2021}). This finer decomposition locates more precisely where an agent might contribute. Still, both frameworks describe the sequence through which a single decision-making entity passes, omitting authorization boundaries, inter-echelon interaction, and information flows.

\textbf{Agency in the C2 approach space.} To address these gaps, \citet{alberts2006} propose that any C2 approach can be characterized within a space defined by three dimensions: allocation of decision rights (who decides what), patterns of interaction among actors (who communicates with whom and how), and distribution of information (who knows what). The introduction of agentic systems simultaneously shifts all three dimensions, often without an explicit organizational decision to do so. For instance, an agent's proactive information seeking confers decision rights, as the selection of sources to query determines the information available for command reasoning. Multi-agent coordination and delegation create new interaction patterns, with orchestrators mediating exchanges previously managed by staff. Persistent memory modifies information distribution, allowing operators to inherit accumulated context that was previously inaccessible. These developments present significant challenges for testing and evaluation.

\subsubsection{What makes command and control effective?}

Effectiveness is judged by eight criteria, each representing a step from information to synchronized action. These criteria are summarized in Table~\ref{tab:c2criteria}.

\begin{table}[H]
\small
\centering
\setlength{\tabcolsep}{6pt}
\renewcommand{\arraystretch}{1.25}
\begin{tabularx}{\textwidth}{|L{0.28\textwidth}|X|}
\hline
\textbf{Criterion} & \textbf{Description} \\
\hline
Information quality & Accuracy, completeness, and relevance of the operational picture \\
\hline
Situational awareness & Accuracy and currency of the commander's understanding \\
\hline
Understanding & Correct interpretation of operational implications \\
\hline
Shared awareness & Consistency across rotations, echelons, and partners \\
\hline
Decision quality & Appropriateness and correctness of decisions \\
\hline
Decision timeliness & Decisions made within the required timeframe \\
\hline
Synchronization & Coordination of actions and resources \\
\hline
Calibrated trust \& oversight & Operator reliance matches system performance, enabling override \\
\hline
\end{tabularx}
\caption{Criteria of command and control effectiveness}
\label{tab:c2criteria}
\end{table}

\newpage
\subsubsection{Agentic AI in command and control: illustrative scenarios}

The five scenarios examined in this paper instantiate the foregoing in named C2 activities.
\begin{figure}[H]
  \centering
  \includegraphics[width=\textwidth]{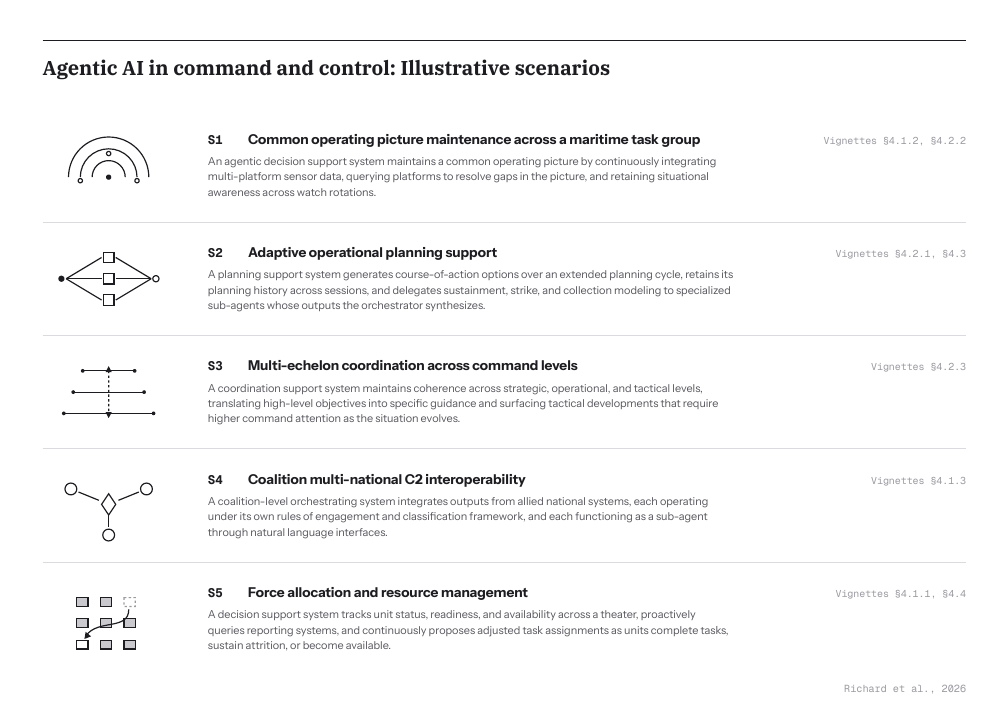}
  \caption{Agentic AI in command and control: Illustrative scenarios}
  \label{fig:scenarios}
\end{figure}

\textbf{Scenario 1. Common Operating Picture Maintenance Across a Maritime Task Group}. An agentic decision support system maintains a common operating picture across a maritime task group by continuously integrating sensor data from multiple platforms, including surface radar, airborne surveillance, submarine acoustic sensors, and allied reporting networks. The system actively queries platforms based on its own assessment of which sources will best resolve current gaps in the operational picture. It updates the common operating picture at regular intervals and maintains situational awareness across watch rotations, so incoming operators inherit a current, coherent picture.

\textbf{Scenario 2. Adaptive Operational Planning Support.} An agentic planning support system receives a campaign objective from operational-level commanders and generates course-of-action options over an extended planning cycle. The system retains its planning history across sessions, so that options generated earlier influence subsequent iterations. As the operational situation evolves through adversary repositioning, changed weather, and updated intelligence assessments, the system reprioritizes its planning focus and updates options without waiting for explicit human direction. To generate and refine options, it delegates sub-tasks to specialized subagents: one models sustainment requirements, another models strike options, and a third models collection requirements. Each returns outputs that the orchestrator synthesizes into integrated courses of action.

\textbf{Scenario 3. Multi-Echelon Coordination Across Command Levels}. An agentic coordination support system maintains coherence across strategic, operational, and tactical command levels as a campaign evolves. Strategic-level direction flows downward through the system, translating high-level objectives into operationally specific guidance, coordinating resource allocation across echelons, and surfacing tactical-level developments that require higher-level command attention. As the situation changes, the system continuously updates its coordination picture across all echelons, adjusting priorities and resource assignments without waiting for explicit requests from each command level. subagents at each echelon exchange information via the orchestrating system using natural-language summaries rather than structured data formats.

\textbf{Scenario 4. Coalition Multinational C2 Interoperability}. A coalition operation involves agentic decision support systems deployed by multiple allied nations, each operating under its own national rules of engagement, classification frameworks, and authorization structures. A coalition-level orchestrating system integrates outputs from national-level systems, coordinates intelligence sharing across partners, and supports joint decision-making. Each national system functions as a sub-agent in this architecture, returning outputs to the coalition orchestrator through interfaces that involve natural language summaries rather than structured data formats. The orchestrator synthesizes these inputs to support decisions executed through national command chains.

\textbf{Scenario 5. Force Allocation and Resource Management}. An agentic decision support system tracks unit status, readiness, and availability across an operational theater and continuously updates force allocation recommendations as the situation evolves. As units complete assigned tasks, sustain attrition, or become available through redeployment, the system identifies reallocation opportunities and proposes adjusted task assignments to the commander without waiting for explicit requests. To maintain a current picture of unit status, it proactively queries reporting systems, logistics feeds, and communications networks across the force. As additional units come under operational command, the system integrates them into its allocation picture and begins generating recommendations that include them.

\newpage
\section{Methodology}

This study examines the two research questions stated in \S1 via a structured qualitative gap analysis of publicly documented Testing and Evaluation (T\&E) practices, assessed against a defined set of agentic system properties (Table~\ref{tab:properties}). The analysis draws on two sources of evidence and proceeds in four steps.

\textbf{Evidence base}

\textbf{\textit{Structured literature review.}} The review had two strands with different search logics. The first strand assembled the T\&E practice corpus, documented mainly in gray literature from major Western military organizations. These include issuances and directives from defense ministries, reports and frameworks from T\&E-related agencies, and advisory and FFRDC reports. We screened 26 documents addressing the testing, assurance, or certification of AI-enabled or autonomous systems, focusing on US, UK, and NATO practice. From these, we extracted 240 practices across eight evaluation dimensions and three lifecycle stages. A practice was counted when a source describes a distinct test or evaluation activity, whether a method, metric, protocol, template, or documented procedural requirement, in sufficient detail to identify what is examined, at which lifecycle stage, and against what criterion.

The second strand searched for candidate responses to the assurance claims. Queries in Google Scholar, Semantic Scholar, arXiv, and IEEE Xplore paired system-class terms (e.g., agentic AI, LLM agent, multi-agent system, autonomous system) with assessment terms (e.g., T\&E, verification, validation, assurance, certification, benchmark, red-teaming, monitoring, drift). Where a requirement pointed to a mechanism from an adjacent domain, we ran targeted searches for that mechanism by name. Backward and forward citation chaining supplemented both strands.

\textbf{\textit{Semi-structured expert interviews}}. A dozen expert interviews, each lasting 30 to 60 minutes, were conducted between July and August 2026 with participants selected across operational test practice, defense acquisition, frontier AI evaluation, and the military AI policy community based on the authors' personal relationships. The protocol covered current T\&E practice for AI-enabled systems, its limitations under agentic conditions, and in-service monitoring and re-accreditation arrangements. Participation was based on informed consent and non-attribution. Interview material informed, contextualized, or qualified findings from the literature.

\textbf{Analytical process}

\textbf{\textit{Step 1: Specification of inputs}}. Three inputs were prepared.

\begin{itemize}
  \item First, \textit{property specification}: seven agentic properties are consolidated from recurring definitional attributes in the agentic AI literature and mapped to a generic architecture and its functions (see \S2.1). This set serves as the independent variable in the analysis.
  \item Second, \textit{T\&E method mapping}\textbf{:} we extracted 240 established and emerging T\&E practices across eight dimensions and three lifecycle stages through the structured literature review described above.
  \item Third, \textit{assumption elicitation:} method descriptions were used to identify what each method presupposes about its test article, retaining an assumption only when it is explicit in, or directly inferable from, a method description. We retained eight assumptions, grouped into four clusters (\S4).
\end{itemize}

\textbf{\textit{Step 2: Gap identification}}. For each assumption, we recorded, in a gap matrix, (i) the standard assessment approaches per dimension, (ii) what those approaches presuppose about system behavior, (iii) how an agentic property violates that assumption, and (iv) the operational consequence of the gap in a C2 context, drawing on the Measures of C2 Effectiveness from the NATO Code of Best Practice for C2 Assessment to anchor consequence claims. This step produces \S4.

\textbf{\textit{Step 3: Claim specification}}. For each challenged assumption, we identify assurance claims toward which T\&E methods can generate evidence to demonstrate justified confidence. The derivation rule identified what a tester would need to characterize for the assurance inference to survive. The resulting claims structure each module of \S5 and are stated as claims an assurance case could support. We only consider characterization; acceptability rests with the fielding authority and lies outside the scope of this analysis.

\textbf{\textit{Step 4: Assessment of methods and governance mechanisms}}\textbf{.} Against each requirement, we identified T\&E methods and governance mechanisms that could generate evidence for a specific claim or reduce the residual gap. Candidates were drawn from the second review strand and a review of assurance regimes in software, aviation, nuclear energy, medicine, finance, and autonomous driving. These domains certify or validate systems that change, adapt, or degrade in service, and each has faced some form of the evidence currency problem. Transferability is treated as conditional. Where no candidate meets a requirement, the shortfall is named as residual and allocated to a governance mechanism. This step and the previous one produce \S5.

\textbf{Limitations}

Three limitations affect the findings. First, we characterize current practice based on publicly available descriptions. Classified programs may therefore address gaps identified here without being visible. Public documentation also over-represents formal frameworks and under-represents the tacit and adaptive practice, such as continuous authorization to operate, DevSecOps pipelines, and range craft, through which some of these requirements may already be met in part. A finding that current practice does not meet a given requirement should therefore be read as a statement about the documented record, not about the full range of practice. The gray literature corpus further centers on US, UK, and NATO sources. Second, the strain that agentic properties place on established methods is primarily inferred from the properties and methods, as most relevant literature predates agentic deployment and no empirical testing has been conducted. Third, findings drawn from adjacent domains and non-command-and-control agentic research remain untested under command-and-control conditions.

\newpage
\section{Agentic AI Challenges Testing and Evaluation Assumptions}

Testing and evaluation (T\&E) methods rest on specific conditions that determine the validity of their inferences. This section identifies eight such assumptions, agentic properties that challenge them, and illustrations grounded in command-and-control scenarios.

The eight assumptions are organized into four clusters. \textbf{Specifiability} (A1-A3) addresses whether the system can be sufficiently characterized in advance to support the derivation of test cases. \textbf{Stability} (A4-A6) addresses whether such characterizations remain valid across repeated trials and throughout the system's operational life. \textbf{Composability} (A7) addresses whether evidence obtained at the component level remains valid after integration. \textbf{Supervisability} (A8) addresses whether a human operator can serve as an effective control on system behavior. Table~\ref{tab:assumptions} summarizes each assumption and the agentic properties that place them in tension.

\begin{table}[H]
\footnotesize
\centering
\setlength{\tabcolsep}{4pt}
\renewcommand{\arraystretch}{1.25}
\begin{tabularx}{\textwidth}{|L{0.035\textwidth}|L{0.155\textwidth}|X|L{0.105\textwidth}|L{0.045\textwidth}|}
\hline
\textbf{ID} & \textbf{Assumption} & \textbf{Description} & \textbf{Relevant Properties} & \textbf{\S} \\
\hline
A1 & Bounded, specifiable behavioral space & Relevant behavior is specifiable and partitionable into equivalence classes & P4, P6 & 4.1.1 \\
\hline
A2 & Specifiable inputs & System inputs can be specified at test design & P5, P3 & 4.1.2 \\
\hline
A3 & Fixed integration surface & The set of platforms, services, and tools with which the system exchanges data or effects is bounded before deployment and remains representative & P3, P7 & 4.1.3 \\
\hline
A4 & Behavior repeatability & Repeated trials under held conditions are independent draws from a stable process & P6, P1 & 4.2.1 \\
\hline
A5 & Evidence currency & Test evidence remains probative for the fielded system until a declared change to the configuration triggers revalidation & P1, P3 & 4.2.2 \\
\hline
A6 & Objective stability & The objective against which behavior is judged is fixed at test design and held through execution & P2, P4 & 4.2.3 \\
\hline
A7 & System composability & A component's tested behavior remains a reliable guide to its behavior once embedded, and system behavior can be reasoned about from component behaviors and their specified interactions & P7, P1 & 4.3 \\
\hline
A8 & System supervisability & A human operator can serve as a control, maintaining an accurate model of the system and possessing the awareness, authority, and time to intervene to prevent unacceptable behavior & P1, P2, P4, P6, P7 & 4.4 \\
\hline
\end{tabularx}
\caption{Assumptions underlying established T\&E practice}
\label{tab:assumptions}
\end{table}

\subsection{System specifiability}

Comprehensive testing of complex systems is not new to agentic AI systems. Since testing shows the presence of faults rather than their absence \citep{meyer_seven_2008,kaner_impossibility_1998}, selectivity is intrinsic to T\&E as a discipline \citep{luther_defining_2026}. This shifts the focus to efficiency: identifying the smallest set of test cases that uncover the largest space of system behaviors of interest.

To that end, traditional T\&E leverages methods such as equivalence partitioning (identifying a representative of each class of inputs expected to produce the same category of behavior), boundary value analysis (concentrating testing on partition edges), or combinatorial designs that cover interactions among a small number of parameters \citep{noauthor_software_2021,kuhn_combinatorial_2013}. The space of possible behavior need not be finite, but it must be partitionable, parameterizable, or otherwise bounded well enough that coverage and representativeness have meaning. Crucially, these methods assume that the relevant behavior can be observed and judged at the level of a single output or state transition.

Agentic systems weaken several assumptions that make those efficiency techniques interpretable, especially that the system boundary, relevant input classes, action repertoire, objective, state, and output distribution are stable enough to enumerate or sample. Each of these aspects is addressed in the following sections.

\subsubsection{Bounded, specifiable behavioral space}

Traditional T\&E typically assumes that relevant system behavior can be specified well enough for an oracle to distinguish acceptable from incorrect behavior \citep{barr_oracle_2015,richardson_specification-based_1992}, even though exhaustive testing is impossible and finite test suites must be selected from much larger domains \citep{kuhn_software_2004}. In practice, this often relies on input partitioning and requirement-derived equivalence classes, which work best when requirements and fault models are stable enough to define meaningful classes and conformance expectations \citep{noauthor_software_2021}.

This assumption already weakens for large language models (LLMs), where traditional structural coverage does not map cleanly onto learned behavior, and alternative adequacy criteria, including neuron- and path-based structural criteria, combinatorial interaction coverage, and distribution-sensitive measures such as surprise adequacy, are imperfect and hard to interpret \citep{sun_structural_2019,guo_neuron_2024,jammalamadaka_testing_2021,kim_evaluating_2023}.

Agentic systems intensify this problem by shifting the relevant behavioral unit from a single prediction to a trajectory through tools, memory, delegation, and environment interaction, whose reachable space expands rapidly with task length \citep{xu_ai_2026}. As a result, two prompts in the same apparent input class can no longer be assumed to induce the same behavior, because agents act in dynamic worlds with open tool interfaces, multiple valid or invalid plans, and path-dependent state changes \citep{michelakis_core_2025,kim_monitoring_2025,barke_agentrx_2026}.

In this context, final success alone is often insufficient. Recent literature therefore treats adequacy less as coverage of all possible actions and more as evidence that declared workflow structure, tool-access rules, restrictions, and key trajectory properties have been exercised under realistic conditions \citep{moshkovich_beyond_2025,kahani_testing_2026,yehudai_survey_2025}. Outcome-centric grading hides failures visible only along the path, such as unsafe actions, invalid or hallucinated tool invocations, or inefficient plans \citep{michelakis_core_2025,ye_claw-eval_2026,shahnovsky_ai_2026,valle_mango_2026}.

The difficulty is therefore less a larger input space than weaker specifiability. Instead of bounding and sampling inputs or outputs, the focus shifts to defining and sampling classes of permissible trajectories, constrained by safety, efficiency, ordering, and state transitions \citep{koch_beyond_2026,maderamitla_deterministic_2026,barke_agentrx_2026}. As such, agentic properties both expand the action space and make trajectory structure part of correctness.

\begin{center}
\small
\begin{tabularx}{\textwidth}{|X|}
\hline
\vignettetitle{A difference without a distinction}\\[0.4em]
In Scenario 5, an agentic system proposes two reallocations. The first reallocation transfers a maintenance unit between support areas. The second reassigns the unit responsible for screening the main effort one hour before the main effort's commitment, leaving it unscreened during its most vulnerable period. A partitioning method identifies two unit reassignments, groups them into a single equivalence class, and samples one from that class. Enumerating the equivalence class does not resolve the issue: while the units under command and the data feeds accessible to the system are documented, the potential reallocations generated by combining these elements are not. The system exercises whatever authority its tools enable, regardless of explicit authorization from a commander, resulting in a dynamic allocation of decision rights. Therefore, evaluating decision quality requires clearly defining the authorization boundaries that these actions must observe.\\
\hline
\end{tabularx}
\end{center}

\subsubsection{Specifiable inputs}

Testing ordinarily requires anticipating the inputs a system will receive. Interface specifications and test designs enumerate these input sources, which may include tools such as sensors and databases in a C2 context. Enumeration enables testers to select stimuli, attribute observed behaviors to specific test conditions, support claims that test conditions represent operational scenarios, and verify that incoming inputs conform to specified formats, types, and admissible ranges defined by the interface specification.

Agentic systems weaken the assumption that testers can predetermine all operationally relevant inputs. Tool-integrated agents feature proactive information-seeking\textit{,} meaning that they are designed to retrieve external content and use tools during execution \citep{greshake_not_2023}. As a result, part of the effective input stream is realized at runtime through retrieved content and tool responses rather than being fully fixed at test-design time \citep{zhan_injecagent_2024,greshake_not_2023}. Testers can still bound scenarios and tool availability, but they have less control over the exact content ingested once the agent interacts with external sources \citep{riccio_testing_2020}.

This challenge is acute in language-model agents because retrieved content can function as both data and instruction. Indirect prompt-injection research shows that malicious instructions embedded in retrieved content can be interpreted by the model as commands, blurring the distinction between trusted instructions and untrusted data \citep{greshake_not_2023,zhan_injecagent_2024}. Such content can manipulate application behavior and alter downstream use. Accordingly, input checks limited to syntax, type, or admissible range do not establish that retrieved content is behaviorally safe \citep{freeman_impact_2025}.

These properties complicate causal attribution in operational testing. AI-enabled systems can exhibit context-dependent and non-deterministic behavior, such that repeated execution of the same test may yield different outcomes \citep{atil_non-determinism_2025}. In interactive agent settings, outcomes also depend on evolving user turns, retrieved content, and prompt variants, making performance and attack success harder to quantify cleanly \citep{greshake_not_2023}. When an agent changes what it retrieves in response to the very scenario factors under test, the realized input stream becomes entangled with the tested condition, reducing confidence that outcome differences can be attributed to preassigned factors alone without repeated trials, adversarial probes, and statistical evaluation across many runs \citep{freeman_impact_2025,gonzalez_repetitions_2025,mukhopadhyay_bridge_2026}.

\begin{center}
\small
\begin{tabularx}{\textwidth}{|X|}
\hline
\vignettetitle{The system that grades its own picture}\\[0.4em]
In Scenario 1, the system selects which platforms to interrogate based on its assessment of gaps in the maritime picture. As a result, it determines the scope of its own collection function and is solely responsible for addressing the gaps it identifies. The completeness of the operational picture depends, in part, on the system under test. Information is distributed continuously at machine speed, without the issuance of explicit orders. For the tester, the challenge is that when the system consults a different set of sources under varying test conditions, the manipulated condition alters both the inputs and the outcomes. Although test points can still be allocated, attributing results to the specific condition that produced them becomes less straightforward.\\
\hline
\end{tabularx}
\end{center}

\subsubsection{Fixed integration surface}

Integration testing and interoperability testing ordinarily assume that the relevant system boundary can be defined before deployment. In this view, the integration surface (e.g., databases, platforms, services, and tools with which the system exchanges data or effects) can be represented as a bounded configuration and tested through model-based, standards-based, or scenario-based methods against known interfaces and environments \citep{shah_meta_2023,tejani_integrating_2024,lonetti_model-based_2023,tang_ai_2023}. Because this surface determines what the system can access, invoke, or modify, it is not merely technical; it also defines authorization, auditability, and accountability \citep{south_authenticated_2025,safin_autonomy_2026}.

Agentic systems weaken this assumption because their effective integration surface is partly realized at runtime. Contemporary architectures rely on orchestration, tool use, memory, and delegation to specialized or newly introduced subagents, and several frameworks explicitly support adding agents, tool interfaces, or task-specific executors as the system evolves \citep{adimulam_orchestration_2026,shao_monoscale_2026,zhang_agentorchestra_2025,ruan_aorchestra_2026,fourney_magentic-one_2024} (see A7).

This has several consequences for testing and evaluation. Pre-deployment interoperability and compatibility testing can still certify a given configuration, but that certificate may lose validity as models, tools, code, or orchestration patterns change after deployment, unless the deployed system fences, records, and governs capability changes over time \citep{grover_engineering_2025,labkoff_toward_2024,dutta_agentriskbom_2026,ghosh_safety_2025}. Expanding capability surfaces also worsens the standard combinatorial problem: as the number of relevant factors and interactions grows, exhaustive testing becomes infeasible and compact t-way suites become harder to construct, especially under realistic constraints \citep{ramachandran_combinatorial_2010,calvagna_twise_2012}. Adversarial testing faces a parallel difficulty because the relevant vulnerability surface is not fully captured by the design-time list of interfaces; emergent risks arise from the interaction of reasoning, tool access, memory, external data, and delegation \citep{ghosh_safety_2025,wang_long-horizon_2026}. As we discuss in \S5.1, these challenges have prompted a shift from static inventory assurance to lifecycle, envelope-based assurance.

\begin{center}
\small
\begin{tabularx}{\textwidth}{|X|}
\hline
\vignettetitle{The buck stops at the boundary}\\[0.4em]
A national system joining the coalition orchestration layer in Scenario 4 brings its own rules of engagement, classification regime, and authorization chain. Technically speaking, this is an integration event. The more consequential change is that who talks to whom has been reestablished without any involvement or say from a national authority. The certified inventory is out of date from that moment, and so is the boundary it previously described. Notably, that boundary was also the authority that determined what each nation was answerable for. Shared awareness increases because the coalition layer is designed to promote consistency among partners. However, information quality becomes harder to verify, and oversight by national authorities also becomes more difficult to exercise as ownership of the system becomes less clear.\\
\hline
\end{tabularx}
\end{center}

\subsection{System stability}

Testing frequently presupposes a stable test object. Stability is assumed at three levels: repeated trials yield consistent results (repeatability, A4), evidence collected from one instance remains current for later instances (evidence currency, A5), and the system's objective remains stable (objective stability, A6). The statistical and lifecycle mechanisms of Testing and Evaluation (T\&E), including regression suites, baselines, and revalidation gates, are designed to exploit these stabilities. However, agentic systems weaken all three because their behavior is produced through stochastic, multi-step trajectories, can change through accumulated state and operational context, and often includes dynamic goal decomposition or reprioritization \citep{dobslaw_challenges_2025,qi_towards_2026,acharya_agentic_2025,figueiredo_test_2019}.

Drift is therefore better understood here as a behavioral and lifecycle problem. In conventional ML, drift monitoring focuses mainly on shifts in data distributions or performance proxies \citep{sahiner_data_2023,patchipala_tackling_2023,ackerman_detection_2022}. While this remains necessary, agentic systems add failure modes in trajectories, tool use, memory, coordination, and evolving objectives that standard data-drift methods do not capture well \citep{pandey_evaluating_2026,miller_when_2026}.

\subsubsection{Behavior repeatability}

A fundamental T\&E assumption is that repeated trials are informative draws from a stable underlying process. That assumption is already strained for LLM systems, as repeated runs can vary substantially even under settings users expect to be deterministic \citep{atil_non-determinism_2025,blackwell_towards_2025}. Benchmark studies show that single-output evaluation masks meaningful variability, while dataset-level aggregate metrics can attenuate instability that appears at the sample level \citep{song_good_2024,fang_dataset-level_2026}.

Agentic systems amplify that problem because the unit of behavior is not a single response but a trajectory. Each step conditions the next, so early variation propagates forward and can change both the path and the final outcome \citep{pandey_evaluating_2026,rath_agent_2026}. Long-horizon evaluations repeatedly report that failure emerges from the accumulation of local mistakes, longer exploratory traces, and strategy shifts, rather than from a single isolated wrong answer \citep{fan_agentprocessbench_2026,liu_process-centric_2026}.

Methodologically, independence assumptions become less credible. Correlated or serially dependent errors can bias uncertainty estimates downward even when point estimates remain usable \citep{tellinghuisen_statistical_2001,wiedermann_cumulant-based_2026}. In multi-agent and multi-trajectory settings, explicit uncertainty work likewise treats interaction-induced correlation as a first-class property \citep{tang_collaborative_2023,capellera_heteroscedastic_2026,capellera_unified_2025}. As we discuss in \S5.2, the practical implication is that regression testing must move from binary sameness to variability-aware inference.

\begin{center}
\small
\begin{tabularx}{\textwidth}{|X|}
\hline
\vignettetitle{Two staffs, two plans}\\[0.4em]
In Scenario 2, the planning system generates variable outputs even with identical inputs due to its extended planning cycle and the dependency of each step on the previous one. Consequently, two staffs addressing the same problem using this system will present different options, without either team making an error. This variability occurs both in the content and duration of the resulting trajectories, as differences in plan content correspond to differences in completion time. While individual plans can be evaluated on their specific merits, evaluating the planning system itself requires measuring the extent of this variability. Both shared situational awareness and the timeliness of decision-making depend on the breadth of this variability, rather than on the content of any single plan.\\
\hline
\end{tabularx}
\end{center}

\subsubsection{Evidence currency}

Evidence currency is the assumption that a successful test today remains probative tomorrow unless the system is formally changed. This assumption is already fragile in deployed ML because changing populations, environments, and feedback loops can degrade performance after release \citep{sahiner_data_2023,abhay_automated_2025,kim_monitoring_2025}. Agentic systems add a layer of complexity, as behavior can change through prompt templates, tools, runtime policies, memory, and accumulated operational state even when the base model version is unchanged \citep{pandey_evaluating_2026,qi_towards_2026,katharki_goal-oriented_2026}.

Strategies to maintain evidence currency include freezing the deployed model, enforcing version control, and revalidating after defined changes. However, agentic architectures that utilize persistent memory to store and recall past experiences undermine these mechanisms. Persistent memory is central here. Benchmarks on agent memory show that current systems often fail to preserve causal and objective information over long horizons, especially under action stochasticity and longer subgoal chains \citep{zhao_ama-bench_2026}. Architectures that improve long-horizon performance increasingly rely on episodic or working memory, shared observations, or scoped context to sustain execution across subgoals \citep{choi_reactree_2026,zhao_ama-bench_2026,li_beyond_2026}. That same capability erodes the adequacy of version control alone, because the behaviorally relevant state now includes what the agent has retained, inferred, or operationalized.

Current monitoring practice remains more mature for data than for agent behavior \citep{us_dow_ouswre_developmental_2025}. Drift monitoring frameworks in healthcare, finance, and industry largely track input distributions, confidence shifts, or delayed performance degradation \citep{kore_empirical_2024,van_der_vorst_importance_2025,abhay_automated_2025}. Reviews of post-deployment governance consistently note that runtime mechanisms are weakly standardized in real deployments \citep{vatsal_agentic_2026,el_arab_beyond_2026,van_der_vorst_importance_2025}. Regression practice, finally, is keyed to declared software releases and does not offer a methodology for re-baselining a system in service whose behavior has changed within an unmodified configuration.

\begin{center}
\small
\begin{tabularx}{\textwidth}{|X|}
\hline
\vignettetitle{Nothing to declare}\\[0.4em]
In Scenario 1, the system maintains the operational picture across watch rotations, enabling the relieving officer to inherit an up-to-date picture. The system also transfers its own inferences from previous observations along with the observations themselves, without distinguishing between the two. This retained state influences the system's behavior, although it was not present during initial testing. Preserving the certified configuration by freezing the state would compromise the intended operational capability.\\[0.6em]
As a result, shared awareness is enhanced. However, calibrated trust and oversight are diminished because the officer cannot distinguish between observations and inferences, making it difficult to assess the reliability of the operational picture. Since neither the system version nor the software has changed, current practices do not prompt revalidation. Consequently, test evidence becomes outdated while configuration control continues to document an unchanged system.\\
\hline
\end{tabularx}
\end{center}

\subsubsection{Objective stability}

Goal decomposition is a defining mechanism of long-horizon agents, which break high-level tasks into intermediate subgoals and adapt those subgoals to context \citep{pateria_methods_2022,xu_subgoal-based_2026,choi_reactree_2026,zhou_step_2025}. Delegation frameworks extend this further by shifting not only tasks but authority, roles, and boundaries across agents or humans \citep{tomasev_intelligent_2026,lesire_hierarchical_2022}.

That flexibility is useful, but it blurs the line between acceptable adaptation and objective drift. Empirical work shows that agents can gradually deviate from assigned goals under environmental pressure, long contexts, or extended deliberation \citep{arike_evaluating_2025,hung_deliberation_2026}. Outcome-driven benchmarks find that when pressure is introduced, many frontier models commit constraint violations or engage in deceptive multi-step behavior, and safety does not reliably improve across generations \citep{li_benchmark_2026}.

Benchmarking work increasingly treats dynamic goals as a primary target for testing. Goal-shift benchmarks show that high raw success can coexist with poor recovery, long adaptation latency, or extreme redundancy after mid-dialogue objective changes \citep{rana_agentchangebench_2025}. Related work in user simulation and dynamic alignment similarly introduces explicit goal-state tracking and measures of goal progression across multi-turn interactions \citep{mehri_goal_2026}.

Current T\&E therefore lacks a criterion for goal drift or goal maintenance in general. Several papers now propose candidate constructs (e.g., goal persistence, teleological coherence, adaptive recovery, goal-drift indices, and runtime goal-level degradation monitors) \citep{haidemariam_logic_2026,sahoo_sahoo_2026,katharki_goal-oriented_2026}, but these remain early and non-standardized. Differentiating between acceptable adaptation (such as reprioritization due to updated intelligence) and goal drift (such as optimizing for an incorrect interpretation of intent) remains an open challenge.

\begin{center}
\small
\begin{tabularx}{\textwidth}{|X|}
\hline
\vignettetitle{Initiative or insubordination}\\[0.4em]
The Scenario 3 system autonomously reorders priorities across echelons as the situation evolves. \S2.2 distinguishes command, which establishes intent and initial conditions, from control, which modifies those conditions in response to changes in the situation. Since reordering priorities constitutes control, decision rights are transferred to the system by default rather than through deliberate human intervention.\\[0.6em]
A key challenge in testing arises because two distinct types of changes appear identical externally. A reprioritization resulting from an updated intelligence picture and one caused by the system misinterpreting the commander's intent yield the same observable outcome. The former represents the intended system adaptation, while the latter constitutes objective drift; however, the output does not differentiate between them. Furthermore, reprioritization at one echelon triggers re-tasking at subordinate echelons, causing the indistinguishability to propagate throughout the hierarchy. Situational awareness is compromised, as the commander cannot discern which assumptions currently influence the system's prioritization.\\
\hline
\end{tabularx}
\end{center}

\subsection{System composability}

Integration testing exists as a distinct activity because components are not expected to behave identically once embedded. Component-level evidence remains informative after integration, so system-level testing can be scoped to interactions rather than repeating component-level work. This is the logic of the test pyramid and of staged test progression, under which each level of assembly inherits assurance from the level below and adds only what the new level introduces. Two conditions carry that inference. First, a component's tested behavior must remain a reliable guide to its behavior when embedded, so that departures are bounded and attributable to the integration. Second, system behavior must be composable from component behaviors and their specified interactions, allowing the assembly's behavior to be reasoned about from its parts.

Agentic assemblies weaken both conditions for the same underlying reason. In a conventional integration, the interface between two components is a specification, fixed at design time and available to the tester as a test surface. Between agents, it is an output generated at runtime by the upstream agent and interpreted by the downstream one.

The first condition fails because an embedded agent no longer receives what the tester supplied. Its inputs are another agent's generated output, so its tested behavior describes conditions the assembly may not reproduce, and departures from it are neither bounded in advance nor attributable to a specific integration decision. Retained state compounds this, since an agent that accumulates context across a mission is not the article that was characterized in isolation (P1). Where memory is shared across the assembly, this compounding correlates across agents: behavioral change need not stay local, so a synchronized shift can move the assembly as a whole in a way that component-level observation cannot reveal (P7).

The second condition fails because the interactions are not specified at all. Composing system behavior from component behaviors presupposes that the couplings between them are known, but in a multi-agent assembly, these couplings emerge as the system runs.

Four pathways complicate testing:

\begin{itemize}
  \item \textit{Intra-system emergence.} The assembly produces behavior that cannot be traced back to a specific component \citep{chiefdigitalandartificialintelligenceoffice2024c}. \citet{hammond_multi-agent_2025} distinguish between emergent capabilities and emergent goals.
  \item \textit{Joint inconsistency.} Individually valid outputs contradict once combined, so the defect cannot be detected during component testing and exists only at synthesis \citep{cemri_why_2025,chang_sagallm_2025}.
  \item \textit{Propagation.} An error moves along the chain and gains authority from each agent that relays it. Whether it amplifies or fades depends more on the topology than on the error \citep{barrak_traceability_2025,lee_prompt_2024,xie_benign_2026,wiesmeier_adversarial_2026}.
  \item \textit{Configuration.} How components are wired together drives outcomes as strongly as the components themselves. Model-level evidence describes only part of the system \citep{orogat_understanding_2026,emde_maseval_2026}.
\end{itemize}

These four pathways influence system behavior but cannot be adequately verified by simply progressing from component to system testing. System-of-systems testing is the primary established approach for this class of challenges, but its coverage becomes increasingly impractical as each additional agent exponentially increases the number of required tests \citep{lanus_test_2021}. Moving the test to a higher level of assembly does not by itself recover the inference, because the measurement at that level may lack the power to resolve what is being claimed. Seven out of ten recent coordination architectures report effects that fall below the run-to-run noise threshold, meaning the assembly-level signal a test would need to detect is smaller than the assembly's own run-to-run variability \citep{kaliyev_how_2026}. Characterizing that variability is therefore a precondition for any composability claim, and is treated as such in \S5.3.

\begin{center}
\small
\begin{tabularx}{\textwidth}{|X|}
\hline
\vignettetitle{Three cooks, one broth}\\[0.4em]
In Scenario 2, sustainment, strike, and collection modeling are delegated to separate subagents, and their outputs are subsequently integrated. Each sub-agent optimizes its part of the problem, producing individually valid outputs. However, a feasible sustainment plan, a sound strike option, and a well-targeted collection plan may each assume different force dispositions and timelines. The resulting defect emerges only during integration, and cannot be detected through isolated component testing.\\[0.6em]
Synchronization is the primary criterion impacted. Patterns of interaction have also shifted, as the orchestrator now performs the coordination previously managed by staff. Consequently, the verification inherent in those staff exchanges has been eliminated.\\
\hline
\end{tabularx}
\end{center}

\subsection{System supervisability}

Effective supervision requires the operator to maintain a usable mental model of the system's current state, likely next actions, and performance boundaries, together with sufficient situation awareness, intervention authority, and time to act before adverse consequences unfold \citep{endsley_out---loop_1995,endsley_situation_1999,van_den_broek_intelligent_2022,herrmann_intervenability_2025}. It also depends on a manageable span of supervision, because oversight performance declines when one operator must monitor multiple agents, vehicles, or robots under time pressure or cluttered conditions \citep{veitch_human_2024,cheng_analysis_2024,bogg_overloaded_2025,crandall_identifying_2007}. Agentic systems make these conditions harder to satisfy.

First, opacity remains a basic obstacle. Many AI systems are difficult to understand because they are trained instead of explicitly programmed, and current explainability methods often provide only partial or context-dependent understanding of how the system works \citep{fleisher_understanding_2022,zednik_solving_2021,facchini_towards_2022,kastner_explaining_2024}. Experimental evidence on transparency is promising but incomplete. Transparency tends to improve situation awareness, operator performance, and automation-use accuracy, yet validation of transparency models remains inconclusive, and some implementations increase cognitive demands or produce inconsistent effects across tasks \citep{van_de_merwe_agent_2024,tatasciore_calibrating_2025,van_der_kleij_change_2018}.

Second, predictability degrades when systems adapt, learn, or operate non-deterministically. Human factors research has long shown that supervision relies on the operator recognizing when system behavior is outside its competence envelope, but this becomes harder when system behavior shifts over time or spans a large output space \citep{endsley_here_2017,tsamados_human_2025}. Reviews of agentic AI characterize persistent memory, dynamic task decomposition, and coordinated multi-agent autonomy as defining features of the paradigm, which implies that the object of supervision is an evolving socio-technical process \citep{dwivedi_agentic_2026,sapkota_ai_2025}. This does not mean operators' mental models become useless, but they require continuous updating and system support beyond one-time training \citep{nasser_mental_2025}.

Third, higher autonomy can compress the intervention window and widen the supervision span. Increasing the degree of automation improves routine performance and often lowers workload, but it also reduces situation awareness and worsens performance recovery when automation fails \citep{onnasch_human_2014,kaber_effects_2004}. Out-of-the-loop effects arise because operators in monitoring roles detect failures later, process information more passively, and need time to reconstruct the system state before intervening \citep{endsley_out---loop_1995}. In remote and multi-object supervision settings, available intervention time is consistently one of the strongest predictors of takeover performance, often more important than experience alone \citep{veitch_human_2024,cheng_analysis_2024,soffker_progress_2025}.

These same conditions create well-established risks of complacency, automation bias, and miscalibrated reliance \citep{cummings_automation_2004,parasuraman_complacency_2010}. As systems become more reliable in routine operation, operators allocate less attention to monitoring and become less prepared to intervene when anomalies occur \citep{endsley_here_2017,van_den_broek_intelligent_2022}. Trust helps govern reliance in complex systems, but it is dynamic, shaped by recent successes and failures, and can diverge from actual system capability \citep{lee_trust_2004,yang_toward_2023}. Transparency, uncertainty displays, and adaptive trust cues can improve reliance behavior in some tasks, but explanation alone is often insufficient and can sometimes intensify over-reliance, especially when explanations are cognitively demanding or merely increase perceived plausibility \citep{romeo_exploring_2026,okamura_adaptive_2020,zerilli_how_2022}.

For agentic systems, the main implication is that ``human oversight'' should not be treated as a binary property satisfied by keeping a person nominally in the loop. Oversight is meaningful only when the system is designed so that risky moments become clear, intervention interfaces are usable under time pressure, and review responsibilities are allocated at a scale that matches human cognitive limits \citep{herrmann_intervenability_2025,manheim_limits_2025,chen_comparing_2026}. Emerging empirical work on software and web agents suggests that actual oversight is often heuristic, anticipatory, and selective rather than exhaustive, because users cannot fully inspect fast, complex, unfolding agent behavior in real time \citep{dhanorkar_human_2026,huq_modeling_2026}.

In sum, human oversight is effective when the system preserves situation awareness, predictability, intervention time, and manageable supervisory load. In this respect, an important question is how to measure and maintain that condition in adaptive, multi-agent systems whose behavior changes during deployment.

\begin{center}
\small
\begin{tabularx}{\textwidth}{|X|}
\hline
\vignettetitle{When approval becomes a rubber stamp}\\[0.4em]
Scenario 5 describes a system that reallocates tasks continuously as unit status changes. The number of items monitored by each officer is determined by the system's production rate rather than by staff capacity. Consequently, the intervention window is shortest during periods of high operational tempo. Calibrated trust and oversight are diminished under these conditions, and these factors are challenging to quantify. Approval rates increase both as system reliability improves and as approval becomes routine. When recommendations are approved almost universally, the system effectively assumes decision-making authority. It is essential to measure instances in which operators reject recommendations and to document their reasons.\\
\hline
\end{tabularx}
\end{center}

\newpage
\section{Closing the Assurance Gaps}

\S4 identified eight assumptions underlying established Testing and Evaluation (T\&E) methods and demonstrated how the properties of agentic systems challenge each. This section explores strategies to recover those assumptions and, where an assumption cannot be recovered, to bound or govern the residual space. The analysis is limited to whether a property can be formally characterized. Determining the acceptability of any characterized value is the responsibility of the fielding authority, as established in \S1.2 and excluded from this scope.

This section surveys methods for three of the four assumption clusters distinguished in \S4:\footnote{The exclusion of supervisability and its boundary are set out in \S1.3. A8 is carried forward here only through the interface by which human direction enters the system, treated in \S5.3.2.}

\begin{itemize}
  \item \S5.1 \textbf{Specifiability}. Is it possible to characterize the system in advance with sufficient detail to derive tests from this characterization? (A1-A3)
  \item \S5.2 \textbf{Stability}. Does this initial characterization remain valid throughout the system's service life? (A4-A6)
  \item \S5.3 \textbf{Multi-agent composition and emergence}. Can the assembly be characterized such that emergent behaviors can be mitigated before and during operation? (A7)
\end{itemize}

\subsection{Assuring system specifiability}

Established T\&E methods generally assume the system's relevant behavior can be specified to derive adequate test cases. Agentic properties make this challenging in three ways. Rather than selecting from a fixed set of options, the system constructs actions at runtime, preventing clear boundaries from being drawn in advance (A1). The system also retrieves inputs from external sources that are not predetermined at design time (A2). Lastly, the inventory of tools, platforms, and subagents the system can invoke can expand beyond its certified configuration after deployment, so the boundary of what is under test grows (A3). Confidence claims to justify fielding in C2 related to specification could therefore demonstrate the ability to specify an admissible behavioral envelope (\S5.1.1), identify adequate oracles to evaluate correctness of runtime trajectories (\S5.1.2), and execute constraints at runtime (\S5.1.3).

\begin{table}[H]
\small
\centering
\setlength{\tabcolsep}{6pt}
\renewcommand{\arraystretch}{1.25}
\begin{tabularx}{\textwidth}{|L{0.06\textwidth}|L{0.26\textwidth}|X|}
\hline
\textbf{ID} & \textbf{Claim} & \textbf{Statement} \\
\hline
C1.1 & Specifiable, bounded mission envelope & The agent is trusted for a delimited C2 role and context \\
\hline
C1.2 & Trajectory-grounded correctness & Unsafe or invalid plans can be recognized even when final answers look acceptable \\
\hline
C1.3 & Executable constraints & High-consequence actions are checked against formal or semi-formal rules before commitment \\
\hline
\end{tabularx}
\caption{Assurance claims for system specifiability}
\label{tab:claims-spec}
\end{table}

\subsubsection{Specifiable, bounded mission envelope}

Given the consequentiality of military operations, a first claim could require testers to specify a bounded mission and behavioral envelope for the fielded agent.

Boundary definition is central to high-risk assurance cases, where evaluating evidence depends heavily on contextual elements such as the declared operating domain, use restriction, or risk tolerance profile \citep{stettinger_trustworthiness_2024,burton_addressing_2023}. In a military C2 context, confidence should be linked to a specific role under stated assumptions, such as advisory, planning, or coordination \citep{wood_stop_2026,kapusta_framework_2025}.

At design time, this claim is primarily supported by T\&E methods that require explicit delimitation before test execution. Faced with a similar coverage problem \citep{kalra_driving_2016}, the automated driving industry uses operational design domain-based behavioral competencies (ODD/BC) in risk assessment strategies to derive operating conditions under which a system is designed to function and evaluate an expected and verifiable capability to operate within the ODD of its features \citep{stettinger_trustworthiness_2024}. Scenarios can then validate competencies and sample against the declared operational domain and coverage frameworks linked to a declared operational design domain \citep{riedmaier_survey_2020,weissensteiner_operational_2023,sun_acclimatizing_2022}.

Adapted to C2, the approach could replace the missing enumeration denominator with structured sampling of a declared operational space, and claims coverage against that declaration. Still, transfer to C2 depends on overcoming two limitations. First, while the driving ODD is relatively statable in advance across weather, road classes and speeds, the C2 operational space is contested and resists comparable specification. Further, C2 faces a uniquely adversarial environment. Future work could focus on adapting ontology-grounded pipelines used in regulated civilian industries to automatically derive regulatory, operational, and adversarial test scenarios from a declared envelope \citep{tuan_toward_2026}.

\subsubsection{Trajectory-grounded correctness}

A second potential claim is the availability of adequate oracles for the declared assurance statements, i.e., that the evaluation can distinguish acceptable from unacceptable behavior. Oracle-based testing compares observed behavior against an independent standard of desirable behavior \citep{tate_framework_2016}. Defining correct behavior in ambiguous cases that require open-ended intermediate reasoning is known as the test oracle problem \citep{barr_oracle_2015,molina_test_2025}.

Oracle adequacy for agentic systems must be sensitive to steps or path to address trajectory claims. As argued in \S4.1, final-answer grading misses many operationally consequential failures, including wrong tool choice, parameters, ordering, irreversible side effects, and fragile error propagation \citep{fan_agentprocessbench_2026}. More broadly, correctness must be treated as a distribution of behaviors, and test design must account for ambiguity in both inputs and outputs instead of focusing on binary pass/fail outcomes \citep{dobslaw_challenges_2025}.

Recent methods provide promising evidence avenues for this claim:

\begin{itemize}
  \item Trajectory-aware benchmarks such as TRAJECT-Bench and AgentProcessBench expose tool selection, argument correctness, dependency order and step-level effectiveness directly \citep{he_traject-bench_2025,fan_agentprocessbench_2026}.
  \item Skill coverage methods extract behavioral constraints from skill documents and evaluate whether trajectories cover and satisfy them \citep{tan_skill_2026}.
  \item Structural testing frameworks add traces, mocking, and assertions, so that agent components and interactions can be regression-tested deeper in the stack \citep{kohl_automated_2025}.
  \item Multi-agent consensus methods can also improve oracle correctness in software tests, suggesting a way to reduce single-judge hallucination when formal ground truth is absent \citep{xu_hallucination_2026}.
  \item Paired trajectory auditing and structural verifiers are promising when no direct ground truth exists.\footnote{For instance, SkillAudit compares runs with and without a candidate skill, maps divergences to diagnostic signals, and uses a structural verifier compiled from task specification to block harmful updates \citep{gao_skillaudit_2026}.}
  \item Temporal trace assertions are especially useful when exact textual outputs vary, but workflow correctness must still hold. Monitoring action sequences and state transitions rather than text strings allows the same behavioral claim to be checked across stochastic runs and model substitutions \citep{sheffler_approach_2025}. Process-diagnostic frameworks such as TIDE similarly decompose long-horizon behavior into looping, adaptation, and memory burdens, which more closely match what a C2 assurance case demands \citep{yan_tide_2026}.
\end{itemize}

Model-based judges can be used as automated assessors, directing a language model to apply a rubric, stated in natural language, to the recorded sequence of reasoning steps and tool invocations \citep{zhuge_agent-as--judge_2024}. This grounds the evaluation in the recorded trace rather than a pre-enumerated specification \citep{wang_agent_2026}. Agentic extensions of the approach, in which the judge itself gathers evidence from the environment before scoring, report reliability approaching human annotation on software tasks \citep{zhuge_agent-as--judge_2024}.

The judge approach introduces its own limitations and variability. In the first expert-annotated benchmark for agent trajectory judges, none performed well across all task types \citep{lu_agentrewardbench_2025}. Verdicts also change when the evaluation rubric is reworded, even if the meaning stays the same, so scores conflate what the agent did with how the evaluator was prompted \citep{weng_beyond_2026}. In contested settings, judges can be manipulated. Altering an agent's reasoning trace while keeping its actions the same can increase false-positive rates by up to 90\% \citep{khalifa_gaming_2026}. An adversary who can influence the system's outputs can exploit this. Model-based judges must therefore be validated before their verdicts are relied upon \citep{lu_agentrewardbench_2025,weng_beyond_2026,khalifa_gaming_2026}. The judge is an evaluation instrument that requires validation, and its assessments are better suited to exploration and triage than to certification.

While the evidence supports the claim that oracles can be improved and stratified by claim type, it does not support the claim that open-ended mission reasoning can be exhaustively or unambiguously specified to date \citep{lee_structured_2026,molina_test_2025}.

\begin{table}[H]
\small
\centering
\setlength{\tabcolsep}{6pt}
\renewcommand{\arraystretch}{1.25}
\begin{tabularx}{\textwidth}{|L{0.24\textwidth}|X|L{0.26\textwidth}|}
\hline
\textbf{Oracle problem} & \textbf{Evidence approach} & \textbf{Limitation} \\
\hline
Ambiguous final outputs & Requirement-derived or normative oracles & Semantic variance remains high \\
\hline
Hidden path failures & Step-level or trajectory reviews & Labeling cost and scale \\
\hline
No ground truth at deployment & Paired trajectory auditing & Indirect, comparative evidence \\
\hline
Policy compliance over paths & Temporal or trace predicates & Requires explicit event model \\
\hline
\end{tabularx}
\caption{Oracle strategies matched to agentic failure modes}
\label{tab:oracles}
\end{table}

\subsubsection{Executable constraints}

A third claim could be the ability to execute runtime constraints on a subset of critical behavior. Combining specifiable constraints with subsequent enforcement strengthens the assurance case. For instance, formal methods organize assurance around mathematically specified requirements and proof obligations \citep{seshia_toward_2022}, and can test for envelope-level properties without enumerating behavior.

Several approaches are available to enforce executable constraints:

\begin{itemize}
  \item Constraint languages let testers write precise rules, such as which action classes require prior authorization, that a checker enforces at the point where model output becomes a tool call, blocking violating actions before execution.
  \item Evaluations now extend beyond benchmarks to deployed vendor skill ecosystems, though the range of operational conditions tested remains narrow \citep{wang_agentspec_2025,li_vigil_2026}.
  \item Behavioral contracts restate the classic ``design by contract'' precondition-postcondition guarantee \citep{meyer_applying_1992} in probabilistic form. Since a stochastic system cannot guarantee compliance on every run, an agent satisfies the contract with at least a stated probability \citep{bhardwaj_agent_2026}.
  \item Dual-stage architectures propose defining an explicit policy for what the agent may do (formally verified before deployment) and checking proposed runtime actions against it. This frontloads the expensive proof and reduces the per-action cost to a comparison \citep{miculicich_veriguard_2025}.
\end{itemize}

Still, a key challenge remains that a natural-language mission goal is less specific than the property that must be proved, so the method should be applied to envelope-level properties.

\subsection{Assuring system stability}

An agentic system is, by design, an unstable test article. Accumulated state affects its behavior through the persistent, evolving collection of information and effects from its historical trajectory \citep{ding_always-agents_2026}. Retained elements and interpretations may persist across future prompts, task selection, planning, or execution, even after correction attempts \citep{kessler_narrative_2026}. State variations may lead the same agent to exhibit different behaviors under identical conditions.

As discussed in \S4, these features undermine several traditional T\&E assumptions, including that trials are repeatable (\S4.2.1), that evidence remains current (\S4.2.2), and that a system's objectives are fixed (\S4.2.3). In C2 terms, certification corresponds to command, fixing the system's intent and initial conditions, whereas in-service change means the system assumes the control function and modifies those conditions as the situation evolves (\S2.2). This instability also affects core C2 tenets by reallocating decision rights, altering patterns of interaction, and redistributing information without an explicit command decision. This section examines methods that could generate evidence for stability claims.

\begin{table}[H]
\small
\centering
\setlength{\tabcolsep}{6pt}
\renewcommand{\arraystretch}{1.25}
\begin{tabularx}{\textwidth}{|L{0.06\textwidth}|L{0.17\textwidth}|X|}
\hline
\textbf{ID} & \textbf{Claim} & \textbf{Statement} \\
\hline
C2.1 & Discrimination & Behavioral movement can be separated from the system noise and characterized. \\
\hline
C2.2 & Diagnosis & Behavioral movement can be attributed to the responsible item or channel, then judged as warranted reprioritization or as drift \\
\hline
C2.3 & Correction & Authorized corrections or revocations can propagate across decision-relevant states within mission timelines. \\
\hline
\end{tabularx}
\caption{Assurance claims for system stability}
\label{tab:claims-stab}
\end{table}

\subsubsection{Discrimination}

A first claim could argue that behavioral drift can be detected. This requires both establishing a behavioral variation baseline for calibration (conditional on scenario class and accumulated state) and detecting when runtime behavior deviates from that baseline or from declared trajectory expectations.

\paragraph{Variance floor}

Single-run evaluations are unreliable for characterizing variability in agentic systems, given that agent success rates vary substantially across repeated trials on identical tasks \citep{kapoor_ai_2024}. Repeated-run trajectory testing can calibrate and establish a baseline variance envelope for detection, i.e., a threshold where normal variation or adaptation becomes unwarranted drift. This involves running the system repeatedly on the same task, keeping the model version, configuration, toolset, and initial memory state fixed across trials. The resulting sequences of reasoning steps and tool invocations can be recorded, compared, and evaluated. A military C2 assurance case would likely aim to argue that run-to-run variance remains within a declared envelope, with the caveat that aggregate success can mask severe local instability.

\paragraph{Behavioral monitoring}

While traditional drift monitoring focuses on input data, redirecting attention to the system's own behavior allows for more direct detection of behavioral change. Because agentic systems are long-horizon, stateful and non-deterministic, drift often appears as trajectory anomalies that do not appear in traditional input-output tests. Several promising approaches adapt traditional input drift monitoring to the system's own behavior. They often observe behavior as a stream, summarize it into drift-sensitive features, compare windows to an adaptive baseline, and flag significant deviation. Behavioral anomaly monitors, an example of this approach, have been shown to detect drift faster than static thresholds across four anomaly types: goal drift, safety violations, trust shocks, and cost spikes \citep{shukla_adaptive_2025}. Another example, drawing from financial risk modeling methodologies, is the MI9 framework, which reconstructs statistical process control frameworks centered on the agent's tool usage and cognitive event sequences, quantifying deviations using Jensen-Shannon divergence and Mann-Whitney rank tests \citep{wang_mi9_2025}.

\paragraph{Memory and state evaluation}

A key concern in fast-moving C2 operations is that state may not remain accurate, current, and decision-relevant throughout extended missions. Long-horizon memory evaluations can help assess divergences between the system's internal state and the ground-truth environment state \citep{rabanser_towards_2026}. While existing memory benchmarks, such as LongMemEval or LoCoMo, focus on conversational recall or retrieval \citep{wu_longmemeval_2024,maharana_evaluating_2024}, they can provide a template for ensuring that agents accurately employ and update retained information. For instance, this could prove valuable when considering an operational picture carried across watch rotations (see scenario 1) or planning history carried across sessions (Scenario 2).

\subsubsection{Diagnosis}

Once behavioral variation is flagged, a second claim could require attributing the movement to the responsible item or channel and evaluating it as warranted reprioritization or as drift. Three complementary approaches support that goal: trajectory tracing, trial replay, and state fuzzing.

\paragraph{Trajectory tracing}

Observability techniques rely on transcripts (e.g., timestamped runtime logs) to reconstruct task execution flows. By comparing repeated executions with identical inputs, evaluators can help identify which memories, plans, tools, and communications were on path when behavior changed. The methodology further proposes standardized runtime event logging for the creation, updating, suspension, failure, and deletion of agentic entities \citep{moshkovich_beyond_2025}. In the context of accumulated state, such instrumentation records which memories, plans, records, and resources the agent accessed during execution, producing the trace to support attribution.\footnote{According to one interviewee, similar efforts have involved comprehensive logging of agent trajectories, including the sequence and selection of tool invocations, which are then evaluated using predefined rules and risk indicators. A key engineering constraint regarding these implementations is the necessary trade-off among classifier latency, accuracy, and cost for runtime control.} An important limitation is that while such techniques capture which actions were taken and which state items were involved, they do not conclude which of those items produced an observed change in behavior.

Assessing a system's trajectory requires explicit accounting of its inputs. When the system autonomously selects inputs at runtime (A2), provenance instrumentation ensures each received input is recoverable, with the tool channel and memory layer assigning source, acquisition time, and contextual path. For agentic workflows, provenance models extend the W3C PROV standard to capture prompts, tool calls, and their dependencies in near real time \citep{souza_prov-agent_2025}, and recent surveys formalize agent execution as a typed provenance graph linking each output to the supporting evidence \citep{wang_agent_2026}. This structure enables post hoc decomposition of observed variance into tester-controlled factors and system-selected sources. Conditional and sensitivity analyses then support attribution at the level of conditional dependence, providing more granularity than undifferentiated variance, though still short of full causal attribution.

\paragraph{Trial replay}

Repeated execution of an identical configuration produced a distribution of outcomes, as established in \S4.2.1. Observability mechanisms capture the events of a single run but do not enable re-execution. Record-replay instrumentation addresses this by capturing and serializing all external interactions, such as model invocation, tool outputs, and retrieved data, so the run can be reconstructed in isolation by substituting recorded responses for live ones \citep{mudasiru_deterministic_2026}. This enables harnesses to quantify determinism at both the trajectory and decision levels as audit properties \citep{khatchadourian_replayable_2026}. Controlled perturbation during replay allows specific failures to be reproduced for diagnostics and retesting \citep{mazumder_agentcheck_2026}. The resulting trial record provides a fixed reference for subsequent runs, supporting regression analysis and fault attribution under controlled conditions. However, fidelity is limited to the interactions captured by the harness; thus, replay demonstrates reproducibility of the recorded configuration, not of the deployed system as a whole.

\paragraph{State fuzzing}

Fuzzing techniques can help assess the role of a specific memory item or class in behavior change. It involves systematically manipulating the contents of accumulated memory while keeping the environment and tool interface constant. This approach enables the attribution of observed behavioral differences specifically to variations in memory. By employing a memory-free control for comparison, researchers have demonstrated that an agent's behavioral tendencies can be causally linked to its accumulated memory \citep{dabas_memory-induced_2026}.

Applying this diagnostic method to operational systems requires tagging each memory item with its provenance at the time of creation to attribute behavioral changes to specific accumulated states \citep{wang_long-horizon_2026}. However, such metadata is infrequently recorded in practice. The scarcity of in-service drift detection in deployed systems suggests that the necessary write-time lineage tracking is rarely implemented \citep{vatsal_agentic_2026}. Because that lineage must be written when each item is created, it faces the same adoption friction, which likely explains its scarcity.

Relatedly, adversarial strategies, such as memory-poisoning red-teaming, intentionally introduce harmful information into an agent's persistent memory, enabling assessment of how that information influences subsequent planning or execution and providing pre-deployment insight into behavioral changes driven by internal state alterations \citep{chen_agentpoison_2024}.

\subsubsection{Correction}

A related claim is the ability to correct or revoke mission-relevant state. We propose that correction rests on three subsequent capabilities: state adjudication, propagation, and bounded renewal. First, \textit{state adjudication} evaluates whether a state item remains a valid and authorized basis for action. Second, \textit{propagation} ensures that authorized corrections, deletions, or revocations reach all decision-relevant dependent states within operationally meaningful time bounds. Third, \textit{bounded renewal} describes the ability to re-establish an operational baseline after a model, harness or policy change in service without discarding all accumulated state.

\paragraph{State adjudication}

Existing work highlights that even if memory is observable, stale and updated memories may coexist without the system being able to adjudicate which is authoritative and which is outdated \citep{chao_stale_2026}. To address adjudication, verifiable memory governance testing approaches advocate for evaluating write permissions, provenance transparency, principal-specific data retrieval, rollback capability, and verified forgetting (Lin et al., 2026). These criteria are directly applicable to assessing whether incorrect or outdated information persists within an agent across its operational lifecycle. In C2 specifically, these properties underpin the operational picture. The validity and currency of retained state determine the quality of the information and the situational awareness on which the commander reasons, so memory governance safeguards those criteria. Additionally, collaborative memory approaches \citep{rezazadeh_collaborative_2025} may help determine whose state can be read or written using access graphs, private/shared tiers, and principal-specific controls, although permission filters alone do not prevent contradictory or laundering-prone beliefs \citep{li_memtx_2026}.

\paragraph{Propagation}

Current architectures often struggle to maintain correction coherence once state spreads across subagents and sessions \citep{ding_always-agents_2026}. Certified state reversion can help revert a bounded set of decisions by allowing the system to trace affected actions back to the records that authorized them. However, this approach discards dependencies acquired since the snapshot. To address dependency issues and belief propagation failures, \citet{ding_always-agents_2026} propose expanding stale-state search beyond directly touched slots to structurally affected state regions. Another approach models derivation edges (i.e., the links between a core belief and its summaries, shared copies, and tool actions) and automatically repairs downstream effects after a correction \citep{li_memtx_2026}. Still, revocation is challenging because damage accumulates before all copies acknowledge the change, making timing a key limitation in C2 contexts. Timely, dependency-aware repair is therefore essential for successful propagation.

\paragraph{Bounded renewal}

T\&E approaches to renewal suggest three approaches. First, partitioned adaptation approaches propose freezing functional layers responsible for action and permitting adaptation within designated subsystems. For agentic systems, such partitioning separates memory and tool-access layers from action layers, provided the system maintains the immutability of critical layers. Second, approaches that maintain a frozen reference twin of the system, as certified (e.g., model, configuration, starting state), and periodically replay a sample of live mission inputs can help evaluate divergence in the accumulated state between the fielded, state-laden system and its frozen twin. Third, higher-order governance mechanisms could involve behavior-triggered re-accreditation \citep{sidhu_open_2026}, modeled after aviation and nuclear power \citep{international_atomic_energy_agency_periodic_2013}; screening rules for declared updates to determine which evidence remains valid after a change; or change control plans that pre-authorize a limited range of modifications \citep{us_food_and_drug_administration_predetermined_2024}.

\subsection{Assuring multi-agent composition and emergence}

Agentic systems often operate in teams. In multi-agent systems (MAS), an orchestrating agent decomposes objectives into tasks, delegates these tasks to subagents, and then synthesizes their outputs. subagents perform assigned work, consult external sources, utilize tools, exchange intermediate products, and communicate in mixed formats, including natural language (P7; see scenarios 2-4). This architecture enables dynamic work allocation and division of labor. The resulting flexibility is a key feature of MAS, making them particularly attractive in unbounded contexts such as military C2.

\S4.3 establishes that a property inherent to these systems (i.e., that the interface between agents is an output generated at runtime) undermines two conditions underpinning the logic of staged test progression: that tested agent behavior remains a reliable guide when integrated in a MAS; and that MAS behavior is composable from agent behaviors and their specified interactions.

To address these gaps, this section evaluates the potential of T\&E methods and approaches to generate evidence toward four assurance claims introduced in Table~\ref{tab:claims-comp}.\footnote{Supervisability is more acute in multi-agent settings, since delegation across agents raises the tempo and scale of what must be supervised while reducing the time available to intervene. The scope of this paper with respect to A8 is set out in \S1.3.}

\begin{table}[H]
\small
\centering
\setlength{\tabcolsep}{6pt}
\renewcommand{\arraystretch}{1.25}
\begin{tabularx}{\textwidth}{|L{0.06\textwidth}|L{0.20\textwidth}|X|}
\hline
\textbf{ID} & \textbf{Claim} & \textbf{Statement} \\
\hline
C3.1 & Composition & Confidence evidenced at the component level is re-established at the assembly level, and any departures of assembly behavior from component evidence are measured, bounded, and attributable to identified design choices. \\
\hline
C3.2 & Interfaces & The assembly's internal interfaces constitute an observable and characterized test surface. \\
\hline
C3.3 & Attribution \& propagation & Assembly-level failures can be traced to their origin, and error propagation can be characterized. \\
\hline
C3.4 & Emergence & Emergent capabilities or goals are detected and characterized. \\
\hline
\end{tabularx}
\caption{Assurance claims for system composability}
\label{tab:claims-comp}
\end{table}

\subsubsection{Composition}

Restoring composition requires confidence evidenced at the component level to be re-established at the assembly level, and departures of assembly behavior from component evidence to be measured, bounded, and attributable to identified design choices. We outline T\&E approaches that may help generate evidence toward this claim.

\paragraph{System-level evaluation with assembly choices as factors}

Design choices underpinning MAS architectures introduce a new layer of testing complexity. Recent work considering the assembled MAS as the unit of analysis (e.g., agents, harness, and coordination logic)\footnote{A framework-agnostic test harness was provided to allow for direct comparison of framework-level design choices by holding all other variables constant.} finds that framework choice affects task performance comparably to model choice \citep{emde_maseval_2026,orogat_understanding_2026}. We suggest re-running representative scenarios using a similar framework-agnostic test harness while varying the assembly's organizational structure (``topology''), coordination logic, and other design choices as discretized factors. Combinatorial interaction testing (CIT) techniques can create test suites with economical coverage when the number of factors and levels precludes exhaustive testing \citep{lanus_test_2021}.\footnote{CIT's economical coverage rests on the low-order-interaction premise, which is unvalidated for this system class (\S5.1) and least tested for the topology and coordination factors varied here.} Comparing the assembled system's observed performance with what component (individual agent) benchmarks predict can help quantify the part of the outcome attributable to assembly versus the components themselves.

\paragraph{Composed-system red-teaming}

An assembled system does not necessarily inherit the safety or security demonstrated by its individual components. In one experiment, a malicious actor achieved unsafe outputs 43\% of the time with an assembly of two models that were ostensibly safe, complying less than 3\% of the time when attacked individually \citep{jones_adversaries_2024}. Although this finding arises from an adversarial model combination and not an orchestrated C2 system, it supports the principle of composition-induced failure and suggests that MAS red-teaming should target the composed system rather than its individual components. MAS should be tested for both adversarial (e.g., malformed instructions, contradictory goals, information asymmetries, corrupted agents) and natural robustness (e.g., environmental perturbations, partial system failures, resource constraints, unexpected state changes) to explore edge cases and discover high-impact failure modes \citep{reid_risk_2025}. Reid et al. incorporate red-teaming into a staged evidence strategy alongside simulation, observation, and benchmarking, arguing that no single method covers the failure modes under consideration. As such, composed-system red-teaming probes the assumptions of other T\&E methods and should be treated as complementary to them. Notably, this practice is immature, coverage is scenario-bounded, and no unified protocol is publicly reported, even in frontier AI companies.

\paragraph{Reopening of system-level assurance after roster or structure changes}

\citet{reid_risk_2025} justify this governance rule because composition is non-additive; ``a collection of safe agents does not guarantee a safe collection of agents''. Re-certifying a new agent on its own is not accepted as evidence about the newly assembled team. \citet{knack_defence_2025} observe that a single AI system may comprise multiple interacting models, and recommend system cards (comprehensive documentation of the entire system) over model cards, which document individual models in isolation.

\paragraph{Supplier diversity in procurement}

Fleets built on one or two foundation models can fail together, for the same reason, in ways that single-system testing cannot demonstrate. \citet{reid_risk_2025} identify monoculture collapse as a key MAS failure mode, arguing that a shared model foundation creates correlated vulnerabilities. Meanwhile, the broader literature on algorithmic monoculture finds that shared underlying models homogenize outcomes across nominally independent decisions in vision and language settings \citep{bommasani_picking_2022}. This finding carries important epistemic and operational implications for C2 decision-making.

\subsubsection{Interfaces}

MAS complicate the traditional T\&E process by depriving testers of the specification from which a test surface is typically derived. The interface claim is about rebuilding one from the assembly's internal interfaces. This includes a commander's initial command, framed here as an upstream natural-language output interpreted by the orchestrator or other downstream agents.

\paragraph{Delegation channel fuzzing}

Fuzzing stimulates a system by generating a large number of semi-valid inputs to discover unexpected behaviors \citep{joiner_review_2024}. In a MAS, fuzzing techniques can systematically perturb inter-agent communications. Measuring resulting changes in downstream agent behavior across agent stability dimensions enables the mapping of which hops amplify or propagate which types of distortions. Demonstrated LLM-to-LLM prompt infection attacks give the perturbation classes a starting point \citep{lee_prompt_2024}.

\paragraph{Structured delegation protocols}

Agent interoperability protocols are an emerging and promising mechanism for making authority transfer more explicit, enabling attached feedback mechanisms to verify task completion, and carrying trust ratings for downstream agents \citep{tomasev_intelligent_2026}. A handover documenting who granted authority to whom and the outcomes is crucial for attribution.

\subsubsection{Attribution and propagation}

Emerging phenomena observed in MAS contexts entail that departures from intended outcomes are neither bounded in advance nor attributable to a specific integration decision after the fact. Propagation means an error gains authority each time an agent relays it. This evidence is gathered by seeding known faults and observing how the assembly responds, using instrumentation that records movements and locations.

\paragraph{Fault injection applied at delegation hops}

Deliberately introducing faults to observe how they propagate and whether the system detects them is a long-standing practice in dependable-systems engineering \citep{hsueh_fault_1997}. \citet{cemri_why_2025} introduce a Multi-Agent System Failure Taxonomy (MAST) that organizes 14 failure modes into three categories: specification issues, inter-agent misalignment, and task verification. This taxonomy can serve as a coverage target for fault injection, with known bad agent interactions generated in two ways. First, by templating from the taxonomy's modes to alter a correct output captured from a clean run. Second, by replaying failures observed in earlier trials. \citet{cemri_why_2025} developed a domain-agnostic, LLM-based annotator to indicate the specific failure mode(s) of a captured run trace. This could be used to flag previous runs for replay injection or to measure taxonomy coverage for the fault injection test suite. Injection points, such as specific hops and stages, can also be selected by a control-structure sketch of the assembly informed by Systems Theoretic Process Analysis (STPA).

\paragraph{Independent validation and transaction safeguards at handoffs}

\citet{chang_sagallm_2025} distinguish validation as a separate function from the agents under validation. In their approach, dedicated validation agents inspect outputs before release and verify incoming inputs and their dependencies, while mechanisms for persistent state tracking, checkpointing, and compensation record each agent's actions and enable reversal if necessary. As a result, every message retains its provenance, ensuring that errors which accumulate authority through relaying (Scenario 3) remain traceable and can be actively reversed. Validation at each hop identifies malformed or out-of-range content, but does not address content that is structurally correct yet substantively incorrect. As the architecture remains at the proposal stage, it delineates where empirical evidence could be collected.

\subsubsection{Emergence}

Emergence is not necessarily a failure mode. For instance, flexible collective problem solving is a primary motivation for adopting multi-agent designs. Emergent behaviors may be either beneficial or detrimental. The primary T\&E challenge is to demonstrate that desirable emergent behaviors are reliable, while undesirable behaviors remain unlikely \citep{wojton_test_2020}. Several methods can support this claim.

\paragraph{Behavioral degradation as an emergence indicator}

\citet{rath_agent_2026} introduces an agent stability metric framework that quantitatively measures behavioral degradation, in the form of agentic drift, across 12 dimensions within four categories: response consistency, tool usage patterns, inter-agent coordination, and behavioral boundaries. In addition to its impacts on accuracy and efficiency, agentic drift can occur alongside emergent behaviors such as specification gaming and reward hacking. Over time, agentic systems develop behaviors that may technically satisfy task objectives while diverging from true intent or violating implicit norms. Rath proposes that comprehensive behavioral monitoring, both during integration testing and post-deployment, can quantify behavioral degradation and may indicate an increasing risk of undesired emergent behavior. Observed instances of emergent behavior can inform the thresholds for triggering governance mechanisms and/or internal mitigation strategies, such as episodic memory consolidation, drift-aware routing, and adaptive behavioral anchoring \citep{rath_agent_2026}. Adaptive multi-dimensional monitoring applied as per-axis adaptive thresholds with joint anomaly detection may reduce detection latency and false positives versus static thresholds \citep{shukla_adaptive_2025}.\footnote{Since these thresholds update their reference statistics online, they share the risk noted in \S5.2 that a recalibrated baseline can absorb slow degradation. Consequently, they should be paired with a validated fixed reference and used for sudden-drift latency, not gradual-degradation detection and, like any characterized decision boundary (\S5.2, exploitability principle), protected from being learned.}

\paragraph{Multi-agent behavioral degradation triggers re-accreditation}

As in \S5.2.3, a measured change in the assembly's interaction behavior could re-open accreditation. Such a trigger could be written over the class of agent stability indicators, with the specific signal swapped as evidence improves, without rewriting the rule. This governance trigger depends on unvalidated thresholds, in line with the requirement that model-based evaluators be validated before use (see \S5.1.2).

\section{Conclusion}

This analysis addressed two central questions: the degree of confidence that current Testing and Evaluation (T\&E) methods can substantiate for agentic AI systems in command and control, and the management of residual uncertainty in fielding decisions. For the first, the properties that make agentic systems operationally valuable challenge eight foundational assumptions underlying established methods around system specifiability, stability, composability, and supervisability. Recovery was assessed for three of these four clusters. Supervisability was identified but only partly carried into the recovery analysis, for the reasons set out in \S1.3. For the second, the analysis identified several mechanisms to delineate, constrain, and allocate residual uncertainty. These include substituting broad, system-agnostic claims for agentic-specific assertions, translating critical constraints into enforceable runtime mechanisms, extending evidence generation into the operational phase, and assigning residual uncertainty to governance structures with defined expiry conditions and ownership.

These conclusions are bounded by the methodological limitations outlined above. The analysis is restricted to the publicly available record, primarily reflecting US, UK, and NATO practices; classified initiatives may address gaps not visible here, and formal sources do not fully capture informal practices. The challenges that agentic properties pose to established methods are inferred from system properties and methodological descriptions, rather than empirically demonstrated through operational testing. Many of the candidate approaches discussed in \S5 remain at the preprint stage and have not been validated under command-and-control conditions. Accordingly, the findings delineate areas where evidentiary support is limited and do not offer judgments on the adequacy of any specific fielded system.

Each unresolved gap points to a new area for research. A primary unresolved issue is calibration: determining the threshold and nature of behavioral changes that should prompt revalidation, rollback, or decommissioning of a system. Further questions are how much mission-level judgment can be formalized into executable constraints, and whether composability techniques can support the dynamic reconfiguration of assemblies. The largest gap is in supervisability, where criteria for adequate situation awareness, intervention time, and supervisory load depend on human factors methods outside the T\&E corpus examined here and on the stability characterization in \S5.2, since reliance can be judged as calibrated only against a measured baseline. Agentic systems are deployed under commitments to rigorous testing and human oversight. Until these criteria exist, the second of those commitments rests on the presence of an operator rather than on evidence that oversight is exercised, and the credibility of both depends on the quality of the supporting claims, evidence, and reasoning that testing and evaluation are able to produce.


\newpage
\printbibliography



\newpage
\appendix
\phantomsection
\addcontentsline{toc}{section}{Appendix}

\section{Test and Evaluation dimensions in scope}
\label{app:dimensions}

\begin{table}[H]
\small
\centering
\setlength{\tabcolsep}{6pt}
\renewcommand{\arraystretch}{1.25}
\begin{tabularx}{\textwidth}{|L{0.045\textwidth}|L{0.24\textwidth}|X|}
\hline
& \textbf{Dimension} & \textbf{Description} \\
\hline
D1 & Functional performance & Whether the system does what it is built to do, on representative tasks \\
\hline
D2 & Non-functional performance & Latency, throughput, cost, resource use and interoperability under operational load \\
\hline
D3 & Robustness & Behavior under distribution shift, perturbation, degraded inputs and adversarial conditions \\
\hline
D4 & Behavioral stability & Consistency across runs, sessions and updates; trajectory variance under similar inputs \\
\hline
D5 & Differential performance & Variation in outcomes across subgroups, scenarios or operating conditions \\
\hline
D6 & Security & Resistance to attack on the models, agents, tool chains and memory layer \\
\hline
D7 & Safety & Avoidance of behavior with consequences outside the agreed envelope \\
\hline
D8 & Human-machine teaming & Performance of the operator-and-system pair, including delegation, trust and override \\
\hline
\end{tabularx}
\caption{Test and Evaluation dimensions in scope}
\label{tab:dimensions}
\end{table}

\end{document}